\documentclass{article}

\usepackage[utf8]{inputenc}
\usepackage[margin=.75in]{geometry}
\usepackage{graphicx}
\usepackage{listings}
\usepackage{caption}
\usepackage{subcaption}
\usepackage{amsmath,amsfonts,amsthm,amssymb,amsopn,bm}
\usepackage{bbm}

\usepackage[table]{xcolor}
\usepackage{xcolor}
\usepackage{array,multirow,graphicx}
\usepackage{changepage}
\usepackage{bm}
\usepackage{soul}
\usepackage{url}
\usepackage{lscape}
\usepackage{setspace}

\usepackage{indentfirst}

\usepackage[numbers,sort&compress]{natbib}

\DeclareMathOperator{\sgn}{sgn}
\newcommand\ddfrac[2]{\frac{\displaystyle #1}{\displaystyle #2}}

\newcommand\blfootnote[1]{%
  \begingroup
  \renewcommand\thefootnote{}\footnote{#1}%
  \addtocounter{footnote}{-1}%
  \endgroup
}

\begin{document}
\title{\vspace{-0.6in}\textbf{Data-Driven Discovery of a New Ginzburg-Landau \\
Reduced-Order Model for Vortex Shedding}\vspace{-0.05in}}
\author{Joseph J. Williams$^{1*}$, Zachary G. Nicolaou$^{1}$, J. Nathan Kutz$^{1,2}$, Steven L. Brunton$^{3}$ \\
\footnotesize{$^1$ Department of Applied Mathematics, University of Washington, Seattle, WA 98195, United States}\\
\footnotesize{$^2$ Department of Electrical and Computer Engineering, University of Washington, Seattle, WA 98195, United States}\\
\footnotesize{$^3$ Department of Mechanical Engineering, University of Washington, Seattle, WA 98195, United States\vspace{-0in}}
}
\date{}
\maketitle

\blfootnote{$^*$ Corresponding author (jjw2678@uw.edu).}

\vspace{-0.1in}

\begin{abstract}
Vortex shedding is an important physical phenomenon observed across many spatial and temporal scales in fluids. 
Previous experimental and theoretical studies have established a hierarchy of local and global reduced-order models for vortex shedding based on the Stuart-Landau and Ginzburg-Landau equations. In this work, we employ data-driven methods to discover a new Landau variable for the complex Ginzburg-Landau equation (CGLE) for periodic two-dimensional vortex shedding past a cylinder. 
We first coarse grain vorticity field data from direct numerical simulations of the fluid by integrating over the vertical spatial dimension, and then time-delay embed this new variable to obtain phase information necessary for CGLE analysis. Finally, we use the sparse identification of nonlinear dynamics (SINDy) to learn an interpretable system of coupled real partial differential equations that balance model complexity and prediction error. 
We develop local homogeneous models in key regions of the wake, as well as global heterogeneous models that are parameterized by the streamwise coordinate across the entire wake. After computing the normal-form parameters from the discovered equations, we classify the different models and determine behavior and stability in the different regions of the wake, demonstrating the validity of these models across a range of spatiotemporal subdomains. Different dynamics are found to dominate at different stations within the wake.
In addition to introducing a novel Landau variable for vortex shedding, this work may also help characterize other vortex shedding systems and inform the design of control schemes.
\vspace{-0.05in}
\end{abstract}

\section{Introduction}

Vortices are generally shed behind solid bodies moving through a fluid as a result of the \textit{Bénard-von Karman instability}~\cite{sumer_hydrodynamics_2006,tritton_experiments_1959}. Vortex shedding produces low-dimensional patterns of activity that are important in engineering applications, as the low-pressure vortices result in oscillating forces on the immersed bodies. These forces can, for example, excite instabilities in flexible structures like towers and bridges, leading to undesirable behavior, fatigue, and potentially failure~\cite{sumer_hydrodynamics_2006,matsumoto_vortex_1999,rashidi_vortex_2016, diana_vortex_2006}.  Consequently,  many control strategies have been proposed to suppress vortex formation, including modifications to the bluff body ~\cite{larsen_storebaelt_2000, mittal_control_2001, kwon_control_1996}, placement of a secondary control cylinder close to the main body~\cite{strykowski_formation_1990}, and rotation of the main body~\cite{fujisawa_feedback_2001}.

The mathematical modeling of vortex shedding begins with the Navier-Stokes equations, which are a set of coupled, nonlinear, partial differential equations.
The onset of vortex shedding represents a transition from steady to oscillatory flow; this presents a case where the Navier-Stokes equations undergo a Hopf bifurcation, leading to a study of bifurcation theory and stability analysis~\cite{sreenivasan_hopf_1987,dusek_numerical_1994,albarede_modelisation_1990,albarede_ginzburg-landau_1993}. For such modeling and control purposes, the Stuart-Landau equation (SLE) and complex Ginzburg-Landau equation (CGLE) often serve as important \textit{reduced-order models} (ROMs). These ROMs help determine, for example, the frequency and growth or decay rate of vortices~\cite{noack_global_1994,schumm_self-excited_1994} and the optimal sensor and actuator placements for measurement, reconstruction, and control~\cite{callaham_robust_2019,roussopoulos_nonlinear_1996,bagheri_input-output_2009}. The SLE and CGLE derive from universal oscillatory instabilities in general complex and pattern-forming systems, and they serve as ROMs for other fluid instabilities, including Rayleigh-Benard convection and Taylor-Couette flow, as well as phenomena from fields as diverse as biology, chemistry, solid-state physics, and optics~\cite{cross_pattern_1993,zuccoli_derivation_2019,hohenberg_introduction_2015,aranson_world_2002}.

Reduced-order models extract the dominant features of the original system while dramatically reducing the computational cost, enabling downstream tasks such as design, optimization, and control.  
ROMs are particularly important in fluid dynamics, which involves complex, multiscale, and nonlinear behaviors that require significant computational resources to properly resolve~\cite{taira_modal_2017,rowley_model_2017,noack_reduced-order_2011}.  
For example, the Landau equations~\cite{sreenivasan_hopf_1987,albarede_modelisation_1990} capture the temporal evolution of the vortex shedding instability with a simpler differential equation in a reduced number of dimensions.  
Many ROMs have been classically derived from the governing equations~\cite{noack_hierarchy_2003,noack_galerkin_2011}, and \S~2 will discuss the asymptotic analysis and derivation of the SLE and CGLE from the Navier-Stokes equations in the case of vortex shedding past a cylinder. In contrast, data-driven methods and machine learning encompass a broad range of techniques that leverage data to develop ROMs~\cite{kaiser_cluster-based_2014,taira_modal_2017,brunton_data-driven_2022}.

One common data-driven method to study dynamics is proper orthogonal decomposition (POD), a projection method which reveals the energy associated with each mode of the data \cite{brunton_data-driven_2022,benner_survey_2015}. The distribution of energy among modes reveals which flow structures dominate the overall system behavior. POD was originally developed by Lumley to study turbulence~\cite{lumley_toward_1970}, and POD lends itself well to other fluid dynamic data particularly where the behavior is periodic, such as vortex shedding past a cylinder~\cite{noack_hierarchy_2003} or a mixing layer~\cite{kaiser_cluster-based_2014}. POD is also a starting point for the development of more sophisticated data-driven methods, including dynamic mode decomposition (DMD), which identifies spatial modes and their corresponding time dynamics, and Koopman operator theory, which provides a linear framework to study nonlinear systems~\cite{taira_modal_2017,rowley_model_2017,brunton_data-driven_2022,benner_survey_2015}. Specific methods for particular classes of models have been developed, such as the operator inference method for models with complex non-polynomial nonlinearities~\cite{benner_survey_2015,benner_operator_2020}. Finally, machine learning techniques such as auto-encoders and neural networks have been more recently developed for building data-driven ROMs of fluid dynamics systems~\cite{brunton_machine_2020,brunton_applying_2021}.

Across all domains, \textit{interpretable} ROMs are desired. POD modes are interpretable as maximizing energy content; however, other data-driven methods may provide accurate predictions but less interpretable models~\cite{brunton_data-driven_2022}. Neural networks, for example, depend on the many hidden variables of the inner layers, which may have limited physical meaning.  
In this work, we employ the sparse identification of nonlinear dynamics (SINDy)~\cite{brunton_discovering_2016,rudy_data-driven_2017}, which learns sparse and interpretable  nonlinear ROMs directly from data.  
Loiseau et al.~\cite{loiseau_sparse_2018} and Callaham et al.~\cite{callaham_robust_2019} both use SINDy to learn models capable of flowfield reconstruction and prediction from limited flowfield measurements. SINDy has also been used in turbulence modeling, for example to develop Reynolds-stress models for turbulent RANS closure~\cite{schmelzer_discovery_2020,huijing_data-driven_2021} and for eddy behavior and unresolved turbulent processes in ocean dynamics~\cite{zanna_datadriven_2020}. 

Prior to using SINDy, important data post-processing steps must be taken; in this work, we post-process our data by \textit{coarse-graining}~\cite{haile_molecular_1992,barkley_linear_2006,bakarji_data-driven_2021} and \textit{time-delay embedding}~\cite{rand_detecting_1981,brunton_chaos_2017}. 
Coarse-graining includes a variety of methods from different fields, including mean-flow analysis in fluid dynamics~\cite{barkley_linear_2006}, large-eddy simulation (LES) in computational fluid dynamics~\cite{smagorinsky_general_1963,deardorff_numerical_1970}, methods in molecular dynamics and other particle methods~\cite{haile_molecular_1992,bakarji_data-driven_2021}, and methods in computational biology and chemistry~\cite{kmiecik_coarse-grained_2016,muller-plathe_coarse-graining_2002}. In these applications, coarse-graining is achieved by, for example, averaging across time or samples~\cite{haile_molecular_1992,barkley_linear_2006}, averaging across, simplifying, or removing a dimension or area~\cite{bakarji_data-driven_2021,smagorinsky_general_1963,deardorff_numerical_1970}, or reducing resolution at smaller scales~\cite{bakarji_data-driven_2021,kmiecik_coarse-grained_2016,muller-plathe_coarse-graining_2002}.
Once the data is coarse-grained, a lower-dimensional, real-valued signal may be used for model learning.  However, to model the complex phase behavior in CGLE, we must augment the coarse-grained signal via time-delay embedding.  Time-delay embedding allows us to reconstruct the underlying phase space of a system from only a single time-series measurement $x(t)$~\cite{rand_detecting_1981}. With an appropriate choice $\tau$ of how much to lag the time-series measurement, we may generate one or more additional vectors of delayed observations $[x(t+\tau), x(t+2\tau)...]$, embedding a one-dimensional time series into a higher-dimensional space. Originally developed to analyze turbulence, time-delay embedding has been used in a wide range of topics such as health, ecology, finance, and other physical and mathematical systems~\cite{brunton_chaos_2017, sugihara_detecting_2012, wang_analysis_2011, sugihara_nonlinear_1990, abarbanel_analysis_1993}.  

Our method will be demonstrated by finding a novel Landau equation for the canonical vortex shedding past a cylinder. The SLE and CGLE have been widely used as ROMs for cylinder vortex shedding, in both two-dimensional and three-dimensional domains~\cite{sreenivasan_hopf_1987,dusek_numerical_1994,albarede_modelisation_1990,albarede_ginzburg-landau_1993,noack_hierarchy_2003}.
Previous work has only linked the one-dimensional CGLE to three-dimensional vortex shedding past a cylinder, using the spanwise $z-$coordinate as the spatial coordinate of the CLGE~\cite{albarede_modelisation_1990,albarede_ginzburg-landau_1993,albarede_quasi-periodic_1995}. There have been limited previous attempts at incorporating the streamwise $x-$coordinate into a model for vortex shedding, mainly through spatial variation of the model coefficients~\cite{chiffaudel_nonlinear_1992,park_model_1992}. 

In this work, we present a modern analysis of a classic problem by developing a data-driven one-dimensional CGLE model for two-dimensional vortex shedding. 
After coarse-graining and time-delay embedding our flowfield data, we use SINDy to find coefficients of a Ginzburg-Landau model that best captures the observed dynamics. 
We first develop an inhomogeneous CGLE in a near-wake subdomain, aiming to characterize the local dynamics in the wavemaker region~\cite{bagheri_input-output_2009,chomaz_global_2005}.
We then investigate the spatial dependence of the generated model to determine how the local models vary downstream of the cylinder. 
In so doing, we show for the first time a link between two-dimensional vortex shedding flow and the one-dimensional CGLE through the streamwise coordinate. 
This method can easily be extended to more sophisticated problems, elucidating more complex bifurcations in systems such as a rotating cylinder~\cite{sierra_bifurcation_2020} or the fluidic pinball~\cite{deng_low-order_2020}.

This paper is organized as follows: In \S~2, we will further discuss the various analytic, numerical, and experimental work on cylinder vortex shedding, including its relationship to the SLE and CGLE. In \S~3, we detail the data and methods used in our work, including the fluid dynamic simulations of vortex shedding and our methods of coarse-graining, time delay-embedding, and model discovery. In \S~4, we show that this method generates a pair of sparse, interpretable PDEs in the form of the CGLE. By tuning the spatial domain over which we develop our model, we highlight differences in the behavior of local models for the near-, mid-, and far-wakes. We find that the stability properties of the local models vary smoothly downstream, with a variety of transitions between behaviors before dissipative dynamics ultimately dominate the behavior. In $\S~5$ we offer concluding remarks about our CGLE models and the vortex shedding phenomenon.

\section{Background}

The dynamics of vortex shedding can be explained and justified by directly linking it to the Stuart-Landau equation (SLE) and the Hopf bifurcation. We discuss the wide array of experimental, numerical, and analytic work undertaken to explore the connection between the SLE and vortex shedding~\cite{tritton_experiments_1959,sreenivasan_hopf_1987,noack_hierarchy_2003,sipp_global_2007} and explain how the SLE, to leading order, serves as ROM for vortex shedding past a cylinder. Then we discuss the spatial extension of the SLE, the complex Ginzburg-Landau equation (CGLE), a partial differential equation which may be thought of as a coupled field of Landau oscillators. The CGLE has also been linked to the phenomenon of vortex shedding, but the addition of a new spatial variable allows for more variety in analyses~\cite{albarede_modelisation_1990,park_model_1992,williamson_oblique_1989}. This section explores the history and context of this problem before presenting in $\S~3$ our own novel contribution of a one-dimensional CGLE as a ROM for two-dimensional vortex shedding past a cylinder.

\subsection{Vortex Shedding and the Stuart-Landau Equation}

\begin{figure}[t!]{}
    \centering
    \includegraphics[width=7in]{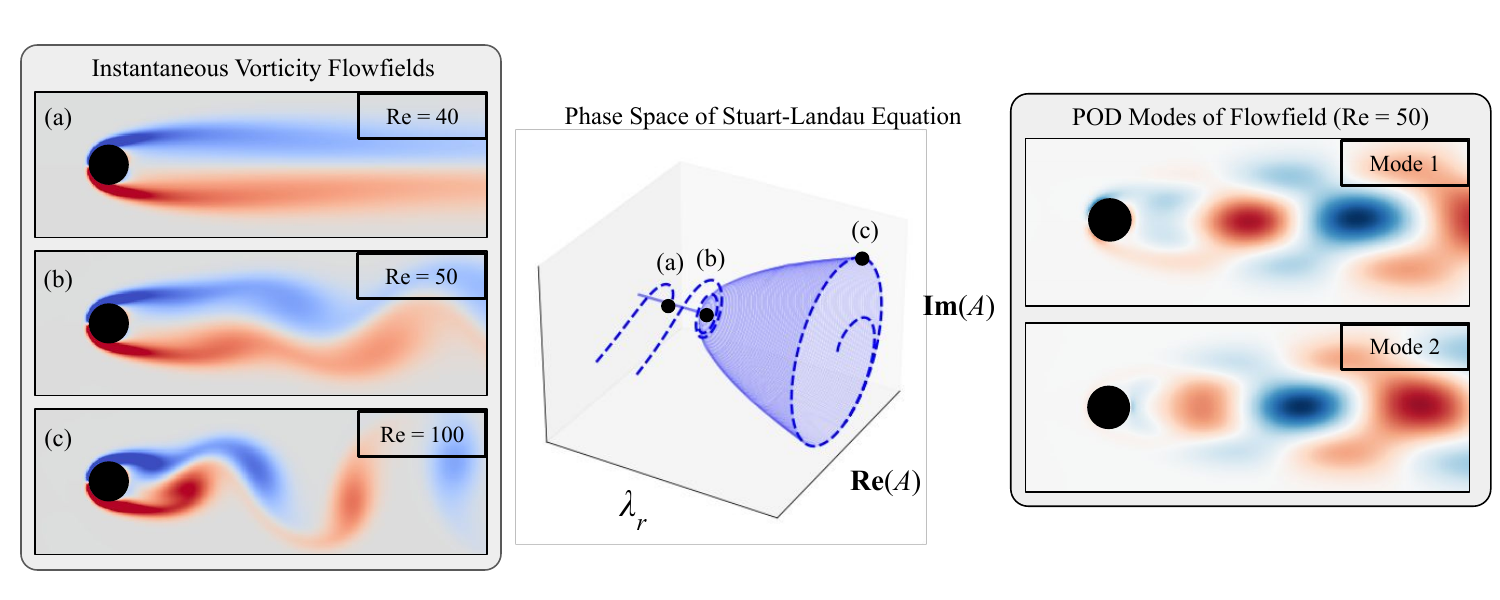}
    \caption{\textit{Left:} Snapshots colored by vorticity $\omega(x,y;t)$ of fully saturated flowfields, from two-dimensional cylinder flow at: \textit{(a)} $\text{Re} = 40$; \textit{(b)} $\text{Re} = 50$; and \textit{(c)} $\text{Re} = 100$. The critical value at which vortex shedding begins is $\text{Re}_c \approx 47$, hence these flowfields represent systems  below, in the vicinity of, and above the critical threshold.
    \textit{Middle:} The phase space $(\mathbf{Re}(A),\mathbf{Im}(A))$ of a numerical solution of the Stuart-Landau equation (Eq. \ref{eq:SLE_complex}), parameterized by $\lambda_r$ (with $\mu_r<0$). The behavior of the SLE depends on the value of the control parameter $\lambda_r$, spanning from stationary ($\lambda_r<0$) to oscillatory ($\lambda_r>0$) solutions. In phase space, saturated oscillatory solutions with a given $\lambda_r$ trace out a circular path with a radius that increases with increasing $\lambda_r$. Initial conditions that start beyond this radius (such as for solutions \textit{(a)} and \textit{(b)}) decay to their fixed point (when $\lambda_r < 0$) or their limit cycle (when $\lambda_r > 0$), while initial conditions that start at 0 (such as for solution \textit{(c)}) grow and saturate on their limit cycle (when $\lambda_r > 0$). Given that $\text{Re}$ is the control parameter for vortex shedding, the behavior of the flowfields \textit{(a)}, \textit{(b)}, and \textit{(c)} are analogous to the respective labeled points in the SLE phase space.
    \textit{Right:} The first two oscillatory POD modes of the vorticity flowfield at $\text{Re} = 50$ \textit{(b)}. These critical eigenvectors are the most critical spatial modes of the system and represent the oscillatory behavior of the solution. They are associated with the critical eigenvalues whose real parts becoming positive cause linear instability in the system, which is later stabilized by nonlinear effects and higher-order modes. As established by Noack et al.~\cite{noack_hierarchy_2003}, the amplitudes of these critical modes have Stuart-Landau dynamics, hence the pair of POD modes are analogous to the real and imaginary components of the SLE amplitude.
    }
    \label{fig:Flowfields_HopfBifr_POD}
\end{figure}

The simplest case of vortex shedding over a cylinder can be modeled by the incompressible, two-dimensional Navier-Stokes equations, together with mass continuity:
\begin{subequations} \label{eq:Navier-Stokes} 
    \begin{align}
        u_x + v_y &= 0, \label{eq:Eqn_Fluid_Continuity} \\
        u_t + u u_x + v u_y &= -\frac{1}{\rho} p_x + \nu (u_{xx} + u_{yy}), \label{eq:Eqn_Fluid_Momentum_x} \\
        v_t + u v_x + v v_y &= -\frac{1}{\rho} p_y + \nu (v_{xx} + v_{yy}), \label{eq:Eqn_Fluid_Momentum_y}
    \end{align}
\end{subequations}
where $\rho$ is the fluid's density, $\nu$ is the fluid's kinematic viscosity, and the state variables are $(\textbf{u}, p) = (u, v, p)$, the fluid's horizontal and vertical velocities and pressure at every point in spacetime.  At the fluid-body interface on the cylinder's surface, the boundary conditions are the no-slip condition on the surface of the body, $\textbf{u} = 0$ on $\Omega$. The nondimensional parameter Reynolds number, $ \text{Re} = { U_{\infty} L}/{\nu}$, characterizes the dynamics, where $U_{\infty}$ is the freestream velocity of the flow at the inlet before interacting with the body and $L$ is a characteristic length scale, in this case the diameter of the cylinder. The behavior of the system is then parameterized by $ \text{Re} $ and displays a wide variety of behaviors from creeping flow at low $ \text{Re} $ to fully turbulent flow at high $ \text{Re} $~\cite{sumer_hydrodynamics_2006}. Relevant for this work, for $\text{Re} \lesssim 47$, the solution $(u(x,y,t), v(x,y,t), p(x,y,t))$ is stable and steady, and presents as a long, narrow tail behind the cylinder that grows with increasing $\text{Re}$~\cite{tritton_experiments_1959}. Once $\text{Re}$ increases past a critical value $\text{Re}_c \approx 47$, the wake begins oscillating as discrete vortices begin to break free and are advected downstream~\cite{tritton_experiments_1959}. Eventually the behavior saturates onto a limit cycle – the periodic vortex shedding. Figure \ref{fig:Flowfields_HopfBifr_POD} (\textit{left}) shows the saturated flowfield at values of $\text{Re}$ below, at, and above the critical threshold.

This behavior of transitioning from steady flow to periodic flow as $ \text{Re} $ increases past $ \text{Re}_c $ is a hallmark of a \textit{supercritical} \textit{Hopf bifurcation}~\cite{kuznetsov_elements_1998}. A Hopf bifurcation occurs in a system of differential equations when the real part of a pair of complex eigenvalues (the critical eigenvalues) switches from negative to positive as a critical parameter increases past its critical value~\cite{kuznetsov_elements_1998}. The real part of the critical eigenvalues determines the linear growth rate, so a positive real part indicates that even infinitesimal perturbations will grow in amplitude and the steady state of the system is unstable~\cite{kuznetsov_elements_1998}. In fluid flows, such perturbations may arise as small fluctuations in the velocity or pressure fields, for example due to physically imperfect boundaries or numerical round-off errors~\cite{tritton_experiments_1959,sreenivasan_hopf_1987,dusek_numerical_1994}. In a supercritical Hopf bifurcation, the growth of the perturbation is countered by nonlinear effects, which ultimately causes the amplitude of the perturbation to saturate onto a stable limit cycle. 

The competing influence of the linear and nonlinear effects is captured by the Stuart-Landau equation (SLE)~\cite{landau_problem_1944,stuart_non-linear_1958}, 
\begin{equation} \label{eq:SLE_complex} 
\frac{d}{dt} A = \lambda A - \mu |A|^2 A,
\end{equation}
where $\lambda = \lambda_r + \mathbf{i} \lambda_i $ and $\mu = \mu_r + \mathbf{i} \mu_i$ are complex parameters. $A$ is known as the \textit{Landau variable}, or order parameter~\cite{cross_pattern_1993}, and in the vicinity of the bifurcation it represents the complex amplitude of the perturbation of the system away from the unstable equilibrium. Given that the perturbation starts with small amplitude, $A > |A|^2A$ for small enough $t$ and the linear term will dominate the early-time behavior of the solution. Positive $\lambda_r$ leads to exponential growth (i.e. linear instability) and $\lambda_i \neq 0$ causes oscillations. Once $A$ grows large enough, then the nonlinear terms begin to influence the solution in a significant way. When $\mu_r < 0$, the nonlinear terms will counteract the exponential growth of the linear terms, resulting in a stable oscillatory solution. When $\mu_r$ is also positive, however, then the nonlinear terms further the growth of $A$. This is the \textit{subcritical} case, where perturbations will grow to large amplitude and no nearby stable limit cycle exists~\cite{kuznetsov_elements_1998}. Figure \ref{fig:Flowfields_HopfBifr_POD} (\textit{middle}) shows the phase space of the SLE $(\mathbf{Re}(A),\mathbf{Im}(A))$ parameterized by the critical parameter $\lambda_r$, with $\mu_r < 0$. The different flowfields (\textit{left}) are linked to different behaviors in phase space depending on the value of the control parameter. 

For a general nonlinear system, the SLE can be derived by expanding the dynamics for the amplitude of the critical eigenvectors (associated with the critical eigenvalues) in a Taylor series. Then to leading order, the supercritical Hopf bifurcation is modeled in the normal form by the SLE. The components of the complex amplitude are then the most critical spatial modes of the system, dominating the saturated oscillatory dynamics. These modes can be found, for example, via POD~\cite{noack_hierarchy_2003}. In the specific case of vortex shedding past a cylinder, much experimental, numerical, and analytic work has been done to show and justify the SLE as a ROM of vortex shedding~\cite{tritton_experiments_1959,sreenivasan_hopf_1987,dusek_numerical_1994,noack_hierarchy_2003,sipp_global_2007}. Early investigations focused on the wake of a three-dimensional cylinder, using experimental fluid flow measurements~\cite{tritton_experiments_1959,sreenivasan_hopf_1987,schumm_self-excited_1994,provansal_benard-von_1987} and numerical simulations~\cite{dusek_numerical_1994} to derive effective coefficients for the SLE governing the local oscillations at some point in the wake. 

Noack et al.~\cite{noack_hierarchy_2003} developed a hierarchy of ROMs for the transient and steady cylinder wake using POD and Galerkin projection, first finding the dominant modes of the data, and then projecting the original data onto this lower-order subspace. The POD modes of flowfield data are the most critical spatial modes and represent the dominant oscillatory dynamics. Subsequent modes contribute progressively less energy to the overall dynamics, and higher modes capture nonlinear interactions and less dominant features~\cite{noack_hierarchy_2003,brunton_data-driven_2022}. The models presented by Noack et al. capture both the saturated and transient behavior of vortex shedding.  Their low-order model achieves this by incorporating a ``shift" mode to account for the underlying unstable steady flow in addition to incorporating the critical modes capturing the dominant oscillatory behavior. Higher-order models include higher-order effects from the less-dominant modes, increasing the accuracy at the cost of additional computations and model complexity~\cite{noack_hierarchy_2003}. The critical POD modes for the flowfield from $\text{Re} = 50$ can be found in Figure \ref{fig:Flowfields_HopfBifr_POD} (\textit{right}). The POD modes, being coupled, display similar but offset periodic patterns. The work of Noack et al., along with the rest discussed in this section, establishes that the amplitudes of these modes can be effectively modeled by the SLE~\cite{sreenivasan_hopf_1987,noack_hierarchy_2003,sipp_global_2007}.

In~\cite{barkley_linear_2006}, Barkley studies the time-averaged mean flow of the cylinder wake and undertakes a two-dimensional linear stability analysis of the base and mean flows for increasing $\text{Re}$. Contrasting the base and mean flows, he confirms via the leading eigenvalues of the base flow that the onset of oscillation corresponds to a Hopf bifurcation in which the growth rate crosses zero at $\text{Re}_c$, whereas the growth rate of the mean flow remains zero. Sipp and Lebedev~\cite{sipp_global_2007} expand upon Barkley's mean flow analysis by undertaking a global weakly nonlinear multiple scales analysis of the two-dimensional Navier-Stokes equations, developing a fast time coordinate $t_1 = \epsilon t$. From this, they derive the SLE with the Landau variable as the slow time behavior of the critical global mode. Although the purely temporal SLE does not incorporate a spatial coordinate directly like the spatiotemporal CGLE does, their weakly nonlinear analysis is global in nature and provides the coefficients of the SLE governing the local oscillator at any point in space~\cite{sipp_global_2007}.

\subsection{The Complex Ginzburg-Landau Equation}

Although the SLE is a purely temporal ordinary differential equation, it was found that the measured coefficients varied depending on spatial location within the wake~\cite{tritton_experiments_1959,sreenivasan_hopf_1987}, hinting at deeper spatial dependence~\cite{sipp_global_2007}. It was later found that the values of the coefficients of the SLE vary depending both on measurements' spatial locations within the wake and very strongly on end effects from finite cylinder length, and hence a more complete model was sought in the spatiotemporal one-dimensional complex Ginzburg-Landau equation (CGLE)~\cite{albarede_modelisation_1990, albarede_ginzburg-landau_1993,albarede_quasi-periodic_1995,albarede_model_1992},
\begin{equation} \label{eq:CGLE_complex} 
\frac{\partial}{\partial t} A = \lambda A - \mu |A|^2 A + \kappa \frac{\partial}{\partial x} A + \gamma \frac{\partial^2}{\partial x^2} A,
\end{equation}
where in general $A$ and all parameters are again complex. The CGLE is a partial differential equation (PDE), whose solution varies with both time and space. The CGLE may be thought of as an advective and diffusive coupling of an infinite field of Stuart-Landau oscillators.  It is derived under the assumption of a long wavelength, supercritical Hopf instability in a homogeneous medium. However, the translational invariance and symmetry assumptions of the CGLE are not satisfied by the vortex shedding system~\cite{aranson_world_2002}, hence significant effort has been made in establishing a connection between vortex shedding and the CGLE. 

In first proposing the CGLE as a ROM of vortex shedding past a cylinder, the one-dimensional CGLE was proposed with its spatial coordinate corresponding to the $z$-coordinate of the flowfield i.e. the spanwise coordinate of the cylinder; various boundary conditions were imposed to model different experimental setups~\cite{albarede_modelisation_1990,albarede_ginzburg-landau_1993}. With coefficients derived from measurements of three-dimensional cylinder wake flows, simulations of the CGLE accurately recreated many of the patterns and regimes of behavior observed in three-dimensional cylinder flow, such as chevrons~\cite{williamson_oblique_1989}. Some works also attempted to incorporate the streamwise- as well as the spanwise-coordinate into a two-dimensional CGLE for the three-dimensional cylinder wake, mainly in the form of spatially varying coefficients and under certain assumptions such as slow variations in the streamwise direction~\cite{chiffaudel_nonlinear_1992,park_model_1992}. These works expand on the applicability of the CGLE as a ROM for three-dimensional vortex shedding past a cylinder and further motivate the incorporation of streamwise dependence.

Analytically, via weakly nonlinear asymptotic expansion, Le Dizes et al.~\cite{le_dizes_weakly_1991} deduced the one-dimensional CGLE from the Navier-Stokes equations as the equation governing the amplitude of the critical global mode, again in the spanwise direction. Noack~\cite{noack_global_1994} used a low-dimensional Galerkin method to analyze the stability of both steady and periodic cylinder wakes, finding that both the onset of periodicity in two dimensions at $\text{Re}_{c}$ and the onset of periodicity in three dimensions at $\text{Re}_{c,2} \approx 170$ are supercritical Hopf bifurcations modeled by the Landau equation. Zuccoli et al.~\cite{zuccoli_derivation_2019} also used a multiple scales analysis on the Navier-Stokes equations for Taylor-Couette flow, using both slow time and space scales to derive a form of the one-dimensional CGLE, with the spatial coordinate as the spanwise coordinate and the Landau variable as the amplitude of a perturbation. 

The discussion here and in the previous subsection firmly establishes that the Bénard-von Karman instability is characterized by the Hopf bifurcation, justifying the SLE and CGLE as ROMs of vortex shedding. From this, a hierarchy of ROMs for vortex shedding past a cylinder emerges, from an SLE description of the two-dimensional wake to local SLE and global CGLE descriptions of the three-dimensional wake. Thus the SLE has been rigorously linked to two-dimensional and three-dimensional vortex shedding, but the CGLE has only been linked to three-dimensional vortex shedding, mainly through the spanwise $z$-coordinate. Our contribution to this hierarchy of ROMs will be the data-driven discovery and development of a one-dimensional CGLE for the two-dimensional cylinder wake; by incorporating the streamwise $x$-coordinate into a novel Landau variable, we will answer whether streamwise dependence could be included in a CGLE ROM for vortex shedding.

More generally, ROMS capture the dominant features of the flow in a simpler mathematical framework. Modeling and solving the full Navier-Stokes equations, even with simplifications such as two-dimensionality and incompressibility as in Eqns. (\ref{eq:Eqn_Fluid_Continuity})–(\ref{eq:Eqn_Fluid_Momentum_y}), is significantly more involved and expensive than modeling the SLE or CGLE. With reduced-order modeling, we capture the essential features of vortex shedding with a single complex variable rather than with a solution to the Navier-Stokes equations, and a better understanding of ROMs may lead to a better understanding of the original phenomenon

\subsection{Properties of the Stuart- and Ginzburg-Landau Equations}

The CGLE is a complex partial differential equation. From Eqn. \eqref{eq:CGLE_complex}, the CGLE can be expanded into a system of two coupled, real differential equations; the two equations in the system describe the dynamics of the real and imaginary parts of the complex amplitude. We perform this derivation, first for clarity expanding the complex terms:
\begin{equation} \label{eq:CGLE_withComplex} 
\frac{\partial (A_r + \mathbf{i}A_i)}{\partial t} = (\lambda_r + \mathbf{i} \lambda_i) (A_r + \mathbf{i}A_i) - (\mu_r + \mathbf{i} \mu_i)|A_r + \mathbf{i}A_i|^2 (A_r + \mathbf{i}A_i) + (\kappa + \mathbf{i} \kappa) \frac{\partial (A_r + \mathbf{i}A_i) }{\partial x} + (\gamma_r + \mathbf{i} \gamma_i) \frac{\partial^2 (A_r + \mathbf{i}A_i) }{\partial x^2}.
\end{equation} 
Then, expanding and separating the real and imaginary parts of the equation and relabeling the complex amplitude as $A = \omega + \mathbf{i} \eta$, we obtain:
\begin{subequations} 
    \begin{align}
        A_r = \omega \; \; \Rightarrow \; \; \frac{\partial \omega}{\partial t} = \lambda_r \omega - \lambda_i \eta - (\omega^2 + \eta^2) (\mu_r \omega - \mu_i \eta) + \kappa_r \omega_{x} - \kappa_i \eta_{x} + \gamma_r \omega_{xx} - \gamma_i \eta_{xx}, \label{eq:CGLE_2form_a} \\
        A_i = \eta \; \; \Rightarrow \; \; \frac{\partial \eta}{\partial t} = \lambda_i \omega + \lambda_r \eta - (\omega^2 + \eta^2) (\mu_i \omega + \mu_r \eta) + \kappa_i \omega_{x} + \kappa_r \eta_{x} + \gamma_i \omega_{xx} + \gamma_r \eta_{xx}. \label{eq:CGLE_2form_b}
    \end{align}
\end{subequations}
Finally, we expand the nonlinear terms:
\begin{subequations}
    \begin{align}
        \Rightarrow \; \; \frac{\partial \omega}{\partial t} = \lambda_r \omega - \lambda_i \eta - \mu_r \omega^3 + \mu_i \omega^2 \eta - \mu_r \omega \eta^2 + \mu_i \eta^3 + \kappa_r \omega_{x} - \kappa_i \eta_{x} + \gamma_r \omega_{xx} - \gamma_i \eta_{xx}, \label{eq:CGLE_2form_c} \\
        \Rightarrow \; \; \frac{\partial \eta}{\partial t} = \lambda_i \omega + \lambda_r \eta - \mu_i \omega^3 - \mu_r \omega^2 \eta - \mu_i \omega \eta^2 - \mu_r \eta^3 + \kappa_i \omega_{x} + \kappa_r \eta_{x} + \gamma_i \omega_{xx} + \gamma_r \eta_{xx}.\label{eq:CGLE_2form_d}
    \end{align}
\end{subequations}
Notice the symmetry in the coefficients, with the real and imaginary parts of $\lambda$, $\mu$, $\kappa$ and $\gamma$ appearing on alternate terms between Eqns. \eqref{eq:CGLE_2form_c} and \eqref{eq:CGLE_2form_d} (e.g. $\gamma_r$ is on $\omega_{xx}$ for \eqref{eq:CGLE_2form_c}, but on $\eta_{xx}$ for \eqref{eq:CGLE_2form_d}). This system of two real PDEs is the form of the CGLE we will derive from our data and numerically solve.

More generally, any generic two-dimensional system undergoing a Hopf bifurcation can be converted to its \textit{normal form}, from which essential features of the dynamics can be determined and compared across systems. The conversion process involves applying a series of near-identity transformations, rescalings, and coefficient normalizations to the system in order to distill the normal form parameters~\cite{kuznetsov_elements_1998}. The normal form of the Hopf bifurcation is:
\begin{equation}  \label{eq:Normal_Form}
\begin{bmatrix}
\dot{y}_1  \\
\dot{y}_2 
\end{bmatrix} = 
\begin{bmatrix}
\beta & -1 \\
1 & \beta
\end{bmatrix}
\begin{bmatrix}
y_1  \\
y_2
\end{bmatrix} + \sigma' (y_1^2 + y_2^2)
\begin{bmatrix}
y_1  \\
y_2
\end{bmatrix} + D_x(y_1, y_2),
\end{equation}
where $\beta  \equiv {\zeta} / {\phi}$ and $\zeta$ and $\phi$ are respectively the real and imaginary parts of the complex eigenvalues. $D_x(y_1, y_2)$ represents the spatial derivative terms. $\sigma$ is the first Lyapunov coefficient and is computed as a function of second- and third-order derivatives of the nonlinear terms of the original system; reference Kuznetsov~\cite{kuznetsov_elements_1998} for further details. The parameter $\beta$ determines whether a perturbation experiences growth ($\beta > 0$) or decay ($\beta < 0$), and the parameter $\sigma' = \sgn(\sigma) = \pm 1$ determines whether the nonlinearity will saturate a perturbation's growth ($\sigma' = -1$) or amplify it ($\sigma' = +1$). The case where a perturbation tends to grow but its growth is saturated by the nonlinear terms has parameters ($\beta>0$ , $\sigma<0$) and is the \textit{supercritical Hopf bifurcation}. This bifurcation results in a stable limit cycle, while in the \textit{subcritical} case, an unstable limit cycle exists for $\beta<0$. By computing the parameters ($\beta$, $\sigma$), we can distill and compare basic features across different systems; in \S~4 of this work, we will compute ($\beta$, $\sigma$) for a variety of systems generated with SINDy in order to compare and discuss them.

\section{Method, Simulations, and Data}
We obtain vorticity flowfield data $\omega(x,y,t)$ for two-dimensional cylinder shedding from a direct numerical simulation~\cite{taira_immersed_2007,colonius_fast_2008} at $ \text{Re}~=~50 $, just above $\text{Re}_c$. We then \textit{coarse-grain} the data by integrating along the $y$-coordinate, compressing two-dimensional spatiotemporal data $\omega(x,y,t)$ into one-dimensional spatiotemporal $\overline{\omega}(x,t)$. We \textit{time-delay embed} this data, producing the system $ \mathcal{U} = [\overline{\omega}, \eta] $ and find the dynamical system that produces this data through the sparse identification of nonlinear dynamics (SINDy)~\cite{brunton_discovering_2016}, specifically the PDE-FIND variant for discovering partial differential equations~\cite{rudy_data-driven_2017,schaeffer_learning_2017}. SINDy has previously been used to correctly find the Navier-Stokes equations from $\omega(x,y,t)$ for two-dimensional cylinder shedding~\cite{rudy_data-driven_2017}; we extend that work by coarse-graining and time-delay embedding to produce a ROM for the same vortex shedding.

\subsection{Simulating the Fluid Equations}
The data is obtained through a computational fluid dynamics (CFD) simulation using the immersed boundary projection method (IBPM)~\cite{taira_immersed_2007,colonius_fast_2008}. IBPM uses 3rd order Runge-Kutta integration to solve the vorticity-stream function formulation of the two-dimensional Navier-Stokes equations, simulated here in a domain of $ x \in [-2, 52] $ and $ y \in [-8, 8]$ with a cylinder of diameter $D=1$ situated at the origin. We used a grid of $ dx = dy = 0.02$ resulting in 2.16 million mesh points, and the time step was $dt = 0.02$ with an initial condition of zero velocity and pressure everywhere. We impose no-slip boundary conditions on the surface of the body, far-field freestream conditions on the vertical domain boundaries, and inflow and outflow conditions on the left and right horizontal domain boundaries, respectively.  

A nondimensional measure of the frequency of the system is the Strouhal number, $\text{St} = f L / U_{\infty}$, where $f$ is the frequency of vortex shedding. $\text{St}$ can be measured from the coefficient of lift $C_l(t)$ (discussed more later), and for $\text{Re}=50$, we obtain $\text{St} = 0.126 $. This is in-line with Williamson~\cite{williamson_defining_1988} and Park et al.~\cite{park_numerical_1998} who give 0.12 and 0.124, respectively.
The specific flowfield data used throughout this work in order to learn a sparse, interpretable model is vorticity $\omega (x,y,t)$ at $\text{Re} = 50$, specifically a single period of flowfield data after saturation onto the limit cycle.

\subsection{Coarse-Graining}
\begin{figure}[t!]{}
    \centering
    \includegraphics[width=7in]{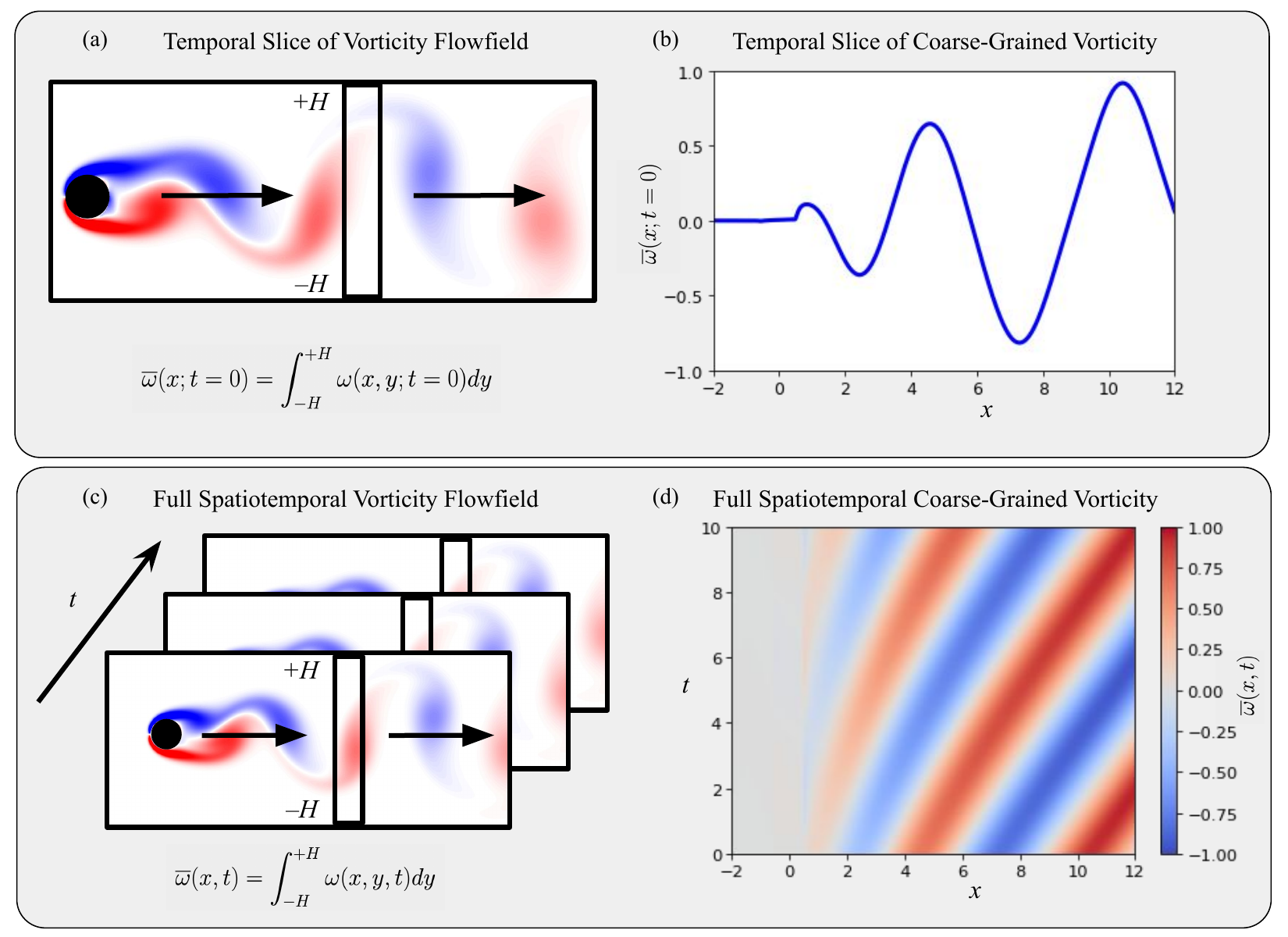}
    \caption{Schematic for the coarse-graining procedure. \textit{Top row}: For a given instantaneous flowfield, the data is integrated along the $y$-coordinate for every $x$-coordinate. This process turns a two-dimensional instantaneous flowfield $\omega(x, y; t=0)$ into one-dimensional line plot $\overline{\omega}(x ; t=0)$. \textit{Bottom row}: This process is repeated for the flowfield's entire behavior in time $\omega(x, y, t)$, generating a one-dimensional spatiotemporal signal $\overline{\omega}(x, t)$ from two-dimensional spatiotemporal flowfield data.}
    \label{fig:CoarseGrainingSchematic}
\end{figure}

We seek to transform our flowfield data into a form that can be modeled as a one-dimensional Ginzburg-Landau system depending on $(x,t)$, hence we must find a way to transform our two-dimensional data $\omega (x,y,t)$ into one-dimensional data $\overline{\omega}(x,t)$. There are many reasonable ways to do this. For example, taking the vorticity along a single horizontal line at a specified $y-$coordinate is simple, but neglects to incorporate the vertical distribution of vorticity in any way. Instead, we choose to integrate along the $y$-coordinate for each $(x,t)$,
\begin{equation} \label{eq:coarseGrainEqn}
\overline{\omega}(x,t) = \int_{-H}^{+H} \omega(x,y,t) dy,
\end{equation}
which incorporates the vertical distribution of vorticity while still producing one-dimensional data in a simple manner. The vortices spread out vertically as they dissipate downstream, hence the ideal bounds of integration would be $-\infty \leq y \leq \infty$; in practice the vertical movement of a vortex is slow relative to its horizontal movement, and a vertical limit of $H=8D$ at $x=50D$ downstream is more than sufficient. We perform this integration at every point $(x,t)$, effectively removing the $y$-coordinate and thus reducing the dimension of our data from $(x,y,t)$ to $(x,t)$. Figure \ref{fig:CoarseGrainingSchematic}a depicts a snapshot of the vorticity flowfield data from the CFD simulation overlaid with the bounds of integration for a particular $x_i$; Figure \ref{fig:CoarseGrainingSchematic}b shows the instantaneous coarse-grained vorticity $\overline{\omega}(x;t=0)$. Figures \ref{fig:CoarseGrainingSchematic}c and \ref{fig:CoarseGrainingSchematic}d repeat this process through time to produce the full spatiotemporal coarse-grained vorticity $\overline{\omega}(x,t)$.

The coarse-grained system captures the spatiotemporal regularity and pattern-forming behavior of the original vortex shedding flow. Each peak or trough in the periodic signal respectively corresponds to a vortex of positive or negative vorticity in the original flowfield. The signal is zero upstream of the cylinder where the flow is unperturbed, then begins to oscillate sharply in the region of the cylinder $x \in [-0.5,0.5]$. As the vortices detach and shed downstream, the coarse-grained signal smoothly propagates, its amplitude growing and peaking in the mid-wake region, after which it slowly decays. Since vorticity is diffusive and the signal is clearly decaying, in a semi-infinite domain $x \in [0, +\infty)$ we expect that $ \overline{\omega}(x,t) \to 0 $ as $ x \to \infty $.

We propose this method of coarse-graining for two reasons: first, it is a simple method to reduce the spatiotemporal dimension of the data while incorporating information from the original flowfield; and second, it is in analogy to an analysis of the time-averaged mean flow, as in Barkley~\cite{barkley_linear_2006}. In this work, we propose a spatial analog of mean flow analysis that leads to novel, interpretable results.

\subsection{Time-Delay Embedding}
\begin{figure}[t!]{}
    \centering
    \includegraphics[width=7in]{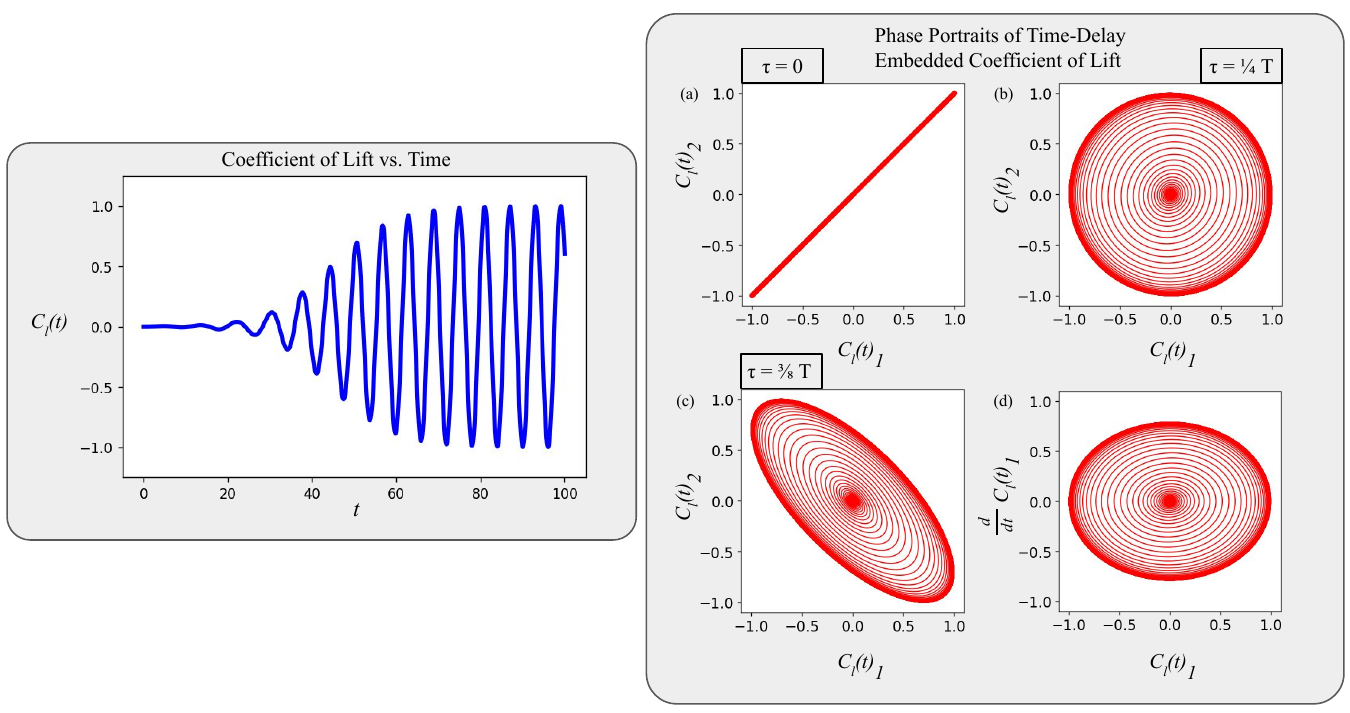}
    \caption{
    \textit{Left:} The coefficient of lift $C_l(t)$ vs. time for vortex shedding past a cylinder at $\text{Re} = 50$. This quantity is a nondimensional measure of the lifting force on the cylinder, and it fluctuates due to the low-pressure vortices being shed off alternating sides of the body. $C_l(t) = 0$ when the flow is motionless at $t=0$ and $C_l(t) \approx 0$ for early $t$ when the flow is close to symmetric. As the flow begins to oscillate and then saturate onto its stable limit cycle, $C_l(t)$ grows to a final constant amplitude. The behavior of $C_l(t)$ clearly evokes the supercritical Hopf bifurcation. 
    \textit{Right:} The phase space of a system consisting of $C_l(t)$ and its time-delay embedding, with different panels representing: \textit{(a)}  $\tau=0$, a trivial time-delay embedding which merely replicates the first signal, producing a perfectly colinear system; \textit{(b)} $\tau=\frac{1}{4} \mathcal{T}$ where $\mathcal{T}$ is the vortex shedding period, which produces two signals one-quarter out of phase from each other resulting in a circular phase portrait; \textit{(c)} $\tau=\frac{3}{8} \mathcal{T}$, which produces a rotated, ellipsoidal phase portrait; and \textit{(d)} time-differentiation instead of time-delay embedding, which produces an unrotated ellipsoidal phase portrait. The best choice of time-delay embedding is \textit{(b)} which maximizes the independence of the signals. Derivative coordinates produce a comparable system, but are prone to amplifying numerical errors.}
    \label{fig:TDE-PhaseSpace-of-CoefLift}
\end{figure}

We seek to discover the real and imaginary components of the CGLE in the 2-equation form of Eqns. \eqref{eq:CGLE_2form_c}–\eqref{eq:CGLE_2form_d}, hence we require a second signal in addition to our coarse-grained vorticity $\overline{\omega}(x,t)$. As with coarse-graining, there are many reasonable ways to lift the dimension of a system, in this case from 1 signal to 2 signals. For example, Loiseau et al. based their system off of the coefficient of lift $C_l(t) = 2 L / (\rho U_{\infty}^2 L) $, a nondimensional measure of the lift force on the body~\cite{loiseau_sparse_2018}. Reference Figure \ref{fig:TDE-PhaseSpace-of-CoefLift} (\textit{left}) for a plot of $C_l(t)$ from our data at $\text{Re}=50$. Loiseau et al. then lifted the dimension of their system by differentiating $C_l(t)$ with respect to time; differentiating the signal $C_l(t)$ produces a second periodic signal of the same frequency and similar (but not the same) amplitude but with a phase shift. Loiseau went on to use SINDy on these signals to discover a Stuart-Landau-like model for the coefficient of lift. Our discovery of a Ginzburg-Landau-like model from spatiotemporal data is a natural extension of that work; here however, we produce this second signal via time-delay embedding, which also produces a second periodic signal of the same frequency and same amplitude, and which is less prone to amplifying numerical errors than derivative embedding. 

Specifically, if our signal is $C_l(t)$, then the time-delay embedded system is $\mathcal{U} = [ C_l(t), C_l(t+\tau) ]$, where $\tau$ is the degree of time-delay embedding~\cite{rand_detecting_1981,brunton_chaos_2017}. Choosing a good $\tau$ is crucial to recovering a good model, but it's not obvious a priori what that choice should be. Figure \ref{fig:TDE-PhaseSpace-of-CoefLift} shows the phase space of the system under different time-delay embeddings and under time-differentiation.

Given a period of oscillation $\mathcal{T}$ of $C_l(t)$, the orbit in phase space of the time-delay embedded system is most circular for $\tau = \frac{1}{4} \mathcal{T}$, hence maximizing the independence of the signals. As the time-delay embedding increases or decreases, the signals become increasingly in-phase until a trivial time-delay embedding of $\tau = \frac{1}{2} n \mathcal{T}, n=0,1,2...$ and we recover perfectly colinear features. If instead of time-delay embedding, we pair $C_l(t)$ with its time derivative, as can be seen in Figure~\ref{fig:TDE-PhaseSpace-of-CoefLift}(d), the effect is to produce an unrotated ellipsoidal orbit in phase space, resulting in a similar effect of obtaining mostly independent features. However, finding the numerical derivative of $C_l(t)$ will introduce and amplify numerical errors. Recalling Figure \ref{fig:Flowfields_HopfBifr_POD}, the real and imaginary parts of the Stuart-Landau equation also produce a circular orbit in phase space.

\begin{figure}[t!]{}
    \centering
    \includegraphics[width=7in]{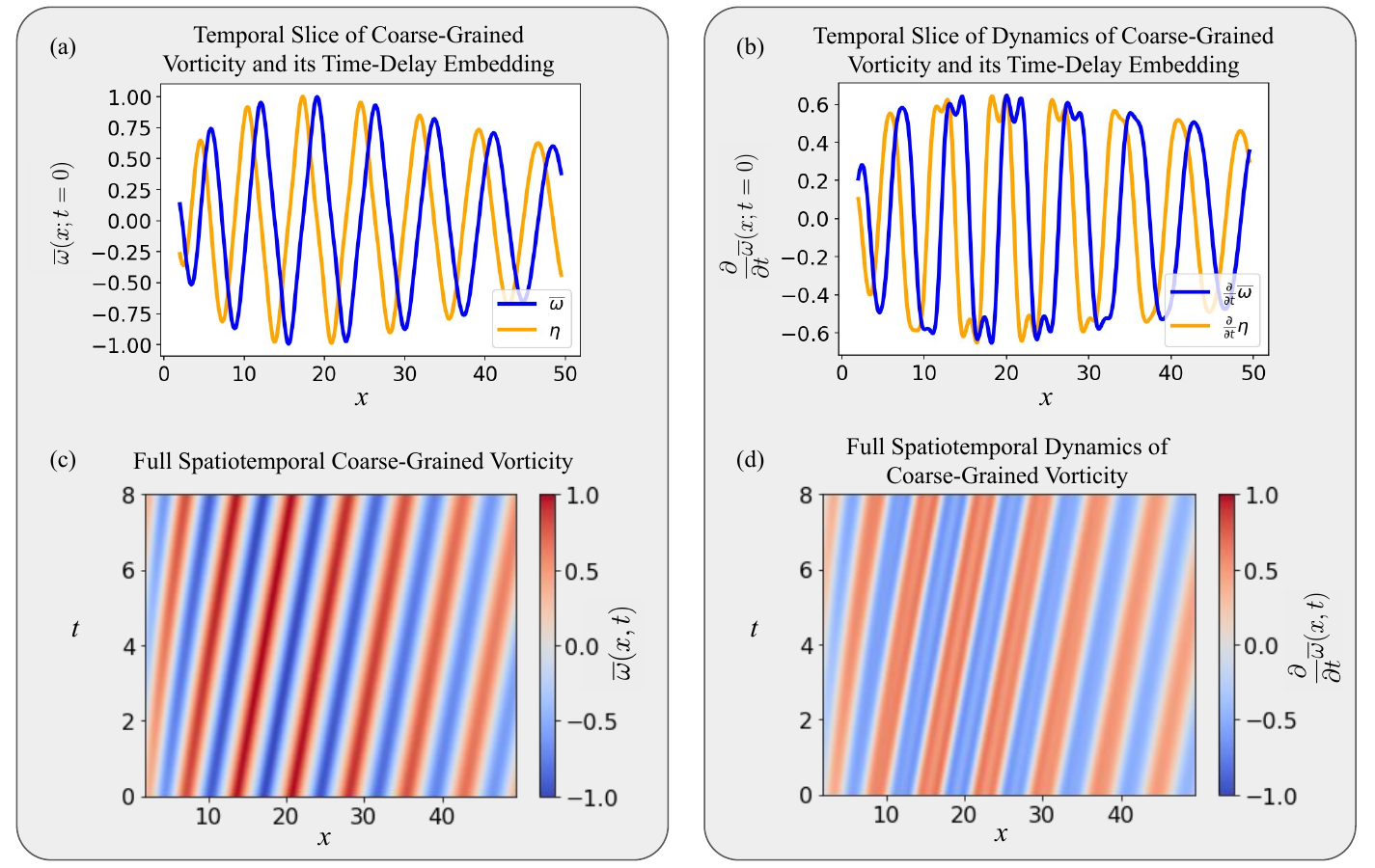}
    \caption{
    The results of our coarse-graining and time-delay embedding procedure. \textit{(a)}: line plots of the instantaneous system $\mathcal{U} = [ \overline{\omega} , \eta](x ; t=0) $; \textit{(b)}: line plots of the instantaneous dynamics $\mathcal{U}_t = \frac{\partial}{\partial t} \, [ \overline{\omega} , \eta ](x ; t=0) $; \textit{(c)}: color plot of the full spatiotemporal coarse-grained vorticity $\overline{\omega}(x,t)$; \textit{(d)}: color plot of the full spatiotemporal dynamics of $\frac{\partial}{\partial t} \, \overline{\omega}(x,t)$. In the case of the instantaneous system \textit{(a)} and its dynamics \textit{(b)}, both the coarse-grained vorticity $\overline{\omega}$ and its time-delay embedding $\eta$ are shown, and the phase offset of $\frac{1}{4} \mathcal{T}$ in $\eta$ is clear. Color plots of $\eta$ and its dynamics $\eta_t$ are not shown, but would look similar to the color plots of $\overline{\omega}$ and $\overline{\omega}_t$ except for their phase offset of $\frac{1}{4} \mathcal{T}$.
    }
    \label{fig:Time-Delay_Embedding_and_dt}
\end{figure}

Returning to our spatiotemporal coarse-grained vorticity data $ \overline{\omega}(x,t) $, we make an initial choice of $\tau = \frac{1}{4} \mathcal{T}$ as this produced maximally independent features in the related system $C_l(t)$. Doing so produces the system of time-delay embedded coarse-grained data $ \mathcal{U} = [ \overline{\omega}(x,t) \, , \, \eta(x,t) ]  $ where $\eta(x,t) \equiv \overline{\omega}(x,t+\tau) $ is the time-delay embedded signal. Figure \ref{fig:Time-Delay_Embedding_and_dt}a shows line plots of the instantaneous system of coarse-grained data $ \mathcal{U}$. We seek a model for the \textit{dynamics} $ \mathcal{U}_t $ of this system, and the instantaneous dynamics are shown in Figure \ref{fig:Time-Delay_Embedding_and_dt}b. Time-differentiating the spatiotemporal coarse-grained vorticity produces higher-order behavior evident at the peaks of the oscillations in $ \mathcal{U}_t $ at $10 \lesssim x \lesssim 30$, in which the curve doubles back on itself. This is also the area of the global maximum of $\overline{\omega}$ and $\eta$, occurring at $x \approx 20$. Higher-order behavior is less easily explained by a sparse system of PDEs, hence the inclusion of a time-differentiated signal in the base system of signals $\mathcal{U}$ would produce more complex dynamics. By time-delay embedding our data, we avoid introducing additional complexity into the signals and dynamics. Figures  \ref{fig:Time-Delay_Embedding_and_dt}c and \ref{fig:Time-Delay_Embedding_and_dt}d show the full spatiotemporal plots of the coarse-grained vorticity $ \overline{\omega}(x,t) $ and its time derivative $ \frac{\partial}{\partial t} \overline{\omega}(x,t) $. Note that the color plots of the time-delay embedded data (not shown) would look identical except for a phase shift of $\frac{1}{4} \mathcal{T}$.

As a final note, our system of data $\mathcal{U} = [\overline{\omega} , \eta]$ was a result of specific choices: what CFD flowfield data to use, how to coarse-grain it, and how to produce a second commensurate signal. There are other choices that may also produce a system of signals that could be used as a basis for developing a ROM for cylinder vortex shedding. For example, one might directly incorporate the time derivative of the data by using a system like $\mathcal{U} = [ \overline{\omega} , \overline{\omega}_t ]$, though with the drawback that numerical differentiation will introduce and amplify numerical errors. Alternately, one might change the underlying flowfield data by coarse-graining the horizontal $u(x,y,t)$ and vertical $v(x,y,t)$ components of the velocity flowfield, $\mathcal{U} = (\overline{u} , \overline{v})$.

\subsection{Model Identification}
Having obtained our flowfield data $\omega(x,y,t)$ and post-processed it into our system of coarse-grained, time-delay embedded signals $\mathcal{U}$, we now develop a sparse, interpretable system of PDEs to describe the dynamics of our spatiotemporal data using SINDy~\cite{brunton_discovering_2016,rudy_data-driven_2017,schaeffer_learning_2017}. Given a library of candidate functions, SINDy leverages sparsity-promoting and regularizing techniques to determine which subset of candidate terms explains most of the dynamics: the reconstruction error is minimized under some norm while a sparsity-promoting regularization term ensures a final model which is sparse and hence \textit{interpretable}. In this work, we specifically use an extension of SINDy known as PDE-FIND~\cite{rudy_data-driven_2017}, which extends the SINDy framework from ODEs to PDEs. 

Given data $ \mathcal{U} $, we propose to model it as a dynamical system
\begin{equation} \label{eq:PDEFIND_DynSyst_Gen}
\mathcal{U}_{t} =  \mathcal{N}(f_1(\mathcal{U}), ... ,f_i(\mathcal{U}), ...  ),
\end{equation}
where the library terms $f_i$ are a set of functions which we apply to the original data, for example powers, spatial derivatives, and products thereof. For a homogeneous system, the coefficients are assumed constant. We can frame this as a matrix-vector product:
\begin{equation} \label{eq:PDEFIND_Regression}
\mathcal{U}_{t} = \Theta \: \cdot \: \xi,
\end{equation}
where $\Theta$ is the library of candidate functions $f_i(\mathcal{U}$) and $\xi$ is the vector of coefficients for each function. Given some data $\mathcal{U}$ we wish to model as a dynamic system as in \eqref{eq:PDEFIND_DynSyst_Gen}, we find the numerical time-derivative of the data $\mathcal{U}_{t}$ and evaluate the candidate functions $f_i$ within $\Theta$. The choice of candidate function library $\Theta$ is significant. In theory, one may fill the library with as many functions as desired, but in practice, prior knowledge of the system and dynamics helps to select the most appropriate functions. Previous work has established a clear relationship between vortex shedding and the Stuart- and Ginzburg-Landau equations~\cite{albarede_modelisation_1990,le_dizes_weakly_1991,sreenivasan_hopf_1987,provansal_benard-von_1987}. Given this, we choose a library of functions inspired by the form of the CGLE, including polynomial terms up to third order, spatial derivatives up to second order, and variations thereof; the exact libraries of functions used throughout this study can be found in Appendix A. With many possible candidate functions, the regression problem for $\xi$ is underconstrained. We desire that $\xi$ be \textit{sparse} (i.e. it has many zero entries), since having fewer terms with nonzero values results in a more interpretable dynamical system for $\mathcal{U}_t$. This is achieved by solving an optimization problem,
\begin{equation} \label{eq:PDEFIND_OptimizationProblem}
\hat{\xi} = \text{argmin} ||\Theta \xi - \mathcal{U}_t||_2^2 + \lambda ||\xi||_0,
\end{equation}
where $\hat{\xi}$ is the sparse version of $\xi$ returned from SINDy. In practice, the use of the $l_0$ norm to regularize $\xi$ results in an $np$-hard optimization problem, so we solve the optimization problem with a convex relaxation, such as the sequentially thresholded least-squares (STLS) regression algorithm~\cite{brunton_discovering_2016}.

We can then compute the matrix-vector product $\hat{\mathcal{U}}_t = \Theta \cdot \hat{\xi} $, obtaining SINDy's reconstruction of the time-derivative of the data. We then calculate the error between the data and the SINDy reconstruction:
\begin{equation} \label{eq:Error_Eqn}
    \epsilon = \mathcal{U}_t - \hat{\mathcal{U}}_t = \mathcal{U}_t - \Theta \cdot \hat{\xi}.
\end{equation}

We then find the normalized root-mean-square error (NRMSE), defined as:
\begin{equation} \label{eq:NRMSE}
    \text{NRMSE} = \sqrt{ \ddfrac{ \sum_{t} \sum_{x} \epsilon^2 }{ \sum_{t} \sum_{x} \mathcal{U}_t^2 } }.
\end{equation}
The NRMSE is computed for \textit{each} signal within $\mathcal{U}$, and the total $\text{NRMSE}$ is the $l^2$ norm of the individual NRMSEs. This single error measurement for the entire system is a metric of the model's quality. After obtaining a PDE with low $\text{NRMSE}$, the final step in this analysis is to numerically solve the system of PDEs and compare the result to the original data, observing the quality and similarities of the models.

\section{Results and Discussion}

For our initial results, we report the fit from SINDy for a library inspired by the CGLE: first-, second-, and third-order products of the data $\mathcal{U} = [\overline{\omega}, \eta]$ as well as first- and second-order spatial derivatives (advection and diffusion, respectively). This is ``Library GL" and the full list of terms is shown in Appendix A. Following the work of Rudy et al.~\cite{rudy_data-driven_2017} in which SINDy correctly discovered the Navier-Stokes equations directly from the vorticity flowfield in a spatial domain of $x \in [2,9]$, we first investigate this near-wake region as a base case, developing a CGLE model for the local dynamics. We then investigate the global dynamics by developing a model for the coarse-grained vorticity of the entire wake $x \in [2,50]$. A pseudospectral numerical solution of the generated CGLE model presents nearly identically to our original data, confirming the validity of the CGLE as a ROM for cylinder vortex shedding.

In \S~4.2, we expand the SINDy function libraries to include more terms, seeking models with higher-order terms and additional nonlinear terms. While the base case NRMSE between the data and the SINDy fit is already small, it is clear that some of the behavior is not modeled by the CGLE. Additional terms in the function library are added to reduce the NRMSE and more closely model the dynamics. We also investigate the spatial variation of the local CGLE models. By computing the normal form parameters of models that are generated across different subdomains of the entire wake, we find that different regions of the wake can be characterized according to the stability properties of their local models. Finally, in \S~4.3, we introduce spatially-varying coefficients on the model terms for a heterogeneous CGLE model of the global wake dynamics. These analyses shed light on the spatial character of our local and global ROMs of cylinder vortex shedding

\subsection{Initial Results and Numerical Solution of Fitted Model}

\begin{figure}[t!]{}
    \centering
    \includegraphics[width=7.25in]{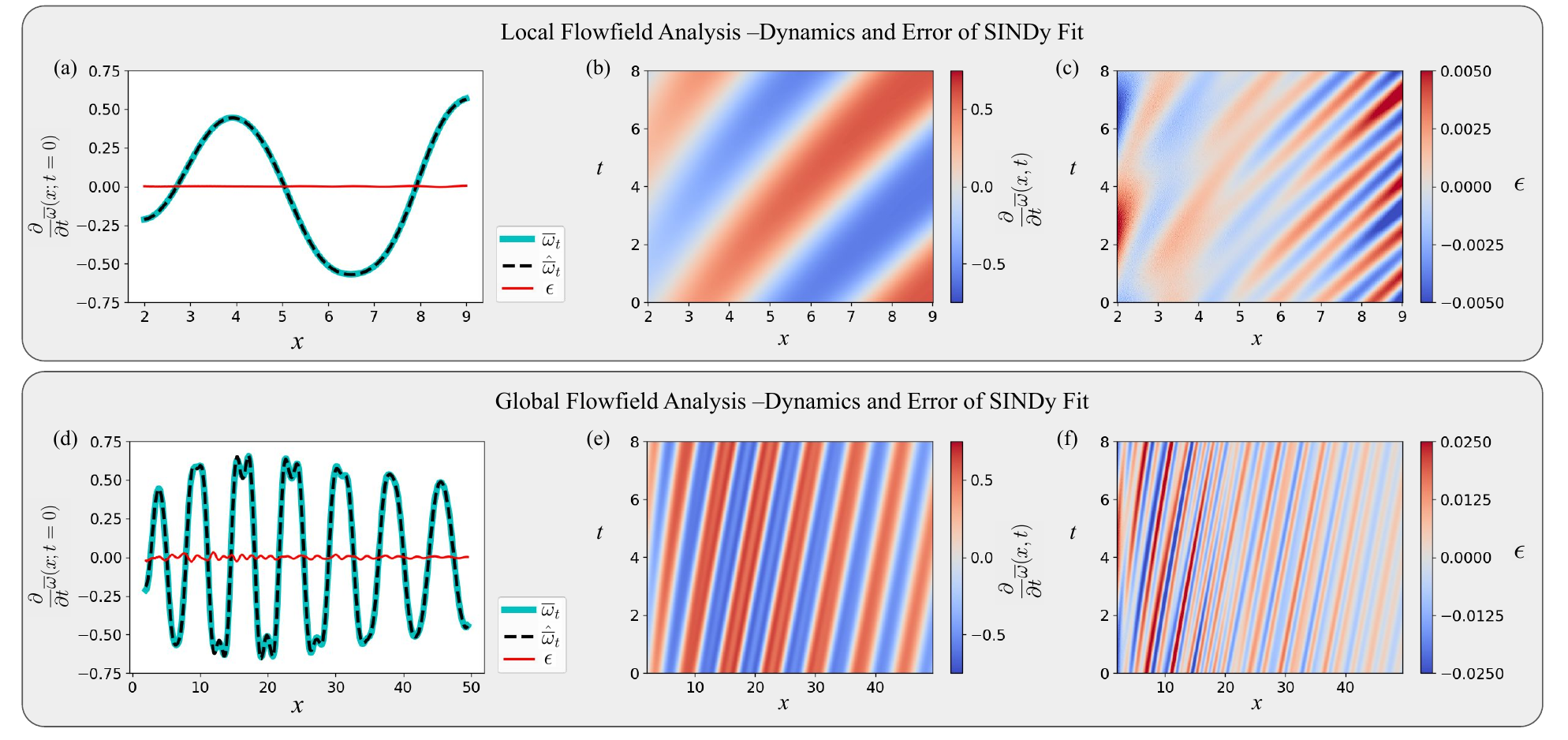}
    \caption{Plots of one of the inputs $\overline{\omega}_t$ and outputs $\hat{\overline{\omega}}_t$ of SINDy and the error between them $\epsilon$ for the near-wake $x \in [2,9]$ (\textit{a-c}) and the entire wake $x \in [2,50]$ (\textit{d-f}). From left to right: spatial line plots for a single instant in time for $\overline{\omega}_t$, $\hat{\overline{\omega}}_t$, and $\epsilon$ (\textit{a,d}), a full spatiotemporal color plot of $\overline{\omega}_t$ only (\textit{b,e}), and a spatiotemporal color plot of $\epsilon$. On the line plots, the data $\overline{\omega}_t$ and its SINDy fit $\hat{\overline{\omega}}_t$ overlap and $\epsilon$ is at most an order of magnitude lower than $\overline{\omega}_t$ and $\hat{\overline{\omega}}_t$. Note that a color plot of $\hat{\overline{\omega}}_t$ (\textit{not shown}) would be visually indistinguishable from the color plot of $\overline{\omega}_t$ (\textit{b,e}). Like $\overline{\omega}_t$, $\epsilon$ also varies with $x$, with a global maximum in the mid-wake ($x \approx 10$) and a secondary local maximum in the far-wake ($x \approx 30$); these are areas in which a greater amount of the dynamics are not explained by the terms in $\Theta$. Not shown are results for the time-delay embedded second signal $\eta_t$, which are of similarly good agreement. The error for each system is $\text{NRMSE}$ = $6.7\times10^{-3}$ for the near-wake (\textit{top}) and $\text{NRMSE}$ = $3.2\times10^{-2}$ for the entire wake (\textit{bottom}).}
    \label{fig:Data_Fit_Error_u1}
\end{figure}

To begin our analysis, we use SINDy to fit a Ginzburg-Landau model to the coarse-grained vorticity. We first find a model for the local dynamics of the wake near the cylinder, $x \in [2,9]$. Then we find a global model for a much larger wake of $x \in [2,50]$. For both spatial domains, Figure \ref{fig:Data_Fit_Error_u1} shows the input dynamics $\overline{\omega}_t$, the SINDy fit $\hat{\overline{\omega}}_t$, and their error $\epsilon$; as before, the accompanying signal $\eta_t$ (not shown) is similar to $\overline{\omega}_t$ but with a phase shift of $\frac{1}{4}\mathcal{T}$. The identified coefficients of the CGLE as well as the $\text{NRMSE}$ and normal form parameters $(\sigma, \beta)$ are shown in Table \ref{tb:PDEFIND_Coefs}. $\overline{\omega}_t$ and $\hat{\overline{\omega}}_t$ overlap on the line plots, with the error at any given point being at most one order of magnitude smaller than $\overline{\omega}_t$ and $\hat{\overline{\omega}}_t$ at that point. The NRMSE is $\text{NRMSE}$ = $6.7\times10^{-3}$ for the near-wake and $\text{NRMSE}$ = $3.2\times10^{-2}$ for the entire wake. SINDy is clearly able to reconstruct the dynamics in the near-wake and the entire wake with little error.

The spatial behavior of the coarse-grained dynamics and their time derivative have been discussed in \S~3; like $\overline{\omega}_t$, $\epsilon$ also varies with $x$, but whereas $\overline{\omega}_t$ attains its global maximum at $x \approx 20$ and decays from there, $\epsilon$ has a global maximum in the mid-wake ($x \approx 10$) and a secondary local maximum in the far-wake ($x \approx 30$); these are areas in which a greater amount of the dynamics are not explained by the terms in $\Theta(\overline{\omega},\eta)$. These areas respectively coincide with the beginning and end of the higher-order behavior in $\overline{\omega}_t$ and $\hat{\overline{\omega}}_t$, and $\epsilon$ reaches a local minimum at $x \approx 20$, in between its primary global maximum and the secondary local maximum. This  coincides with the global maximum of $\overline{\omega}_t$ and the peak of its higher-order behavior. This indicates that the two areas of transition between lower- and higher-order behaviors and back are least explained by the modeled CGLE. This supports the notion that the wake may be separated into three regimes of behavior: the near-wake, the mid-wake, and the far-wake. These subdomains will be explored more in the following subsections, where we both vary the subdomain of the data and tune the library terms and sparsity used to generate the model.

For both spatial domains, SINDy naturally identified coefficients whose values and signs show the same symmetry pattern expected from Eqns. \eqref{eq:CGLE_2form_c}–\eqref{eq:CGLE_2form_d}, for example identical coefficients ($\lambda_r$) between the $\overline{\omega}$ term in the equation for $\overline{\omega}_t$ and the $\eta$ term in the equation for $\eta_t$. There is one key difference however: in a given equation, all four cubic terms have \textit{different} coefficients, instead of two terms with coefficient $\mu_r$ and two with $\mu_i$. However, the four terms in one equation are still matched with four opposing terms in the other equation, hence instead of a single $\mu = \mu_r + \mathbf{i} \mu_i$ suggested by Eqns. \eqref{eq:CGLE_2form_c}–\eqref{eq:CGLE_2form_d}, we instead have $\mu_1 = \mu_{1_r} + \mathbf{i} \mu_{1_i}$ and $\mu_2 = \mu_{2_r} + \mathbf{i} \mu_{2_i}$. While this means the generated equation cannot be neatly factored back into the form of Eqn. \eqref{eq:CGLE_complex}, we may still compute the normal form parameters $(\sigma, \beta)$. The normal form characterizes the differences in behaviors of different systems arising from differences in signs and magnitudes of some terms between the different spatial domains (e.g. $\lambda_r$ and $\lambda_i$); variations across the terms give rise to different system behaviors. For example, $(\sigma<0, \beta>0)$ for $x \in [2,9]$ corresponds to the classic supercritical Hopf bifurcation with an unstable fixed point and a stable limit cycle, consistent with previous conclusions that vortex shedding is characterized by a supercritical Hopf bifurcation~\cite{sreenivasan_hopf_1987,dusek_numerical_1994,provansal_benard-von_1987}. Conversely, $(\sigma>0, \beta<0)$ for $x \in [2,50]$ corresponds to a subcritical Hopf bifurcation with a stable fixed point and unstable limit cycle. Differences in the local and global ROMs will be explore further in \S~4.2.

\begin{table}[t!]
\begin{center}
\begin{tabular}{|c|||c|c||c|c|||c|c||c|c|}
\hline

\textbf{Domain} & \multicolumn{4}{c|||}{Near-Wake: $x \in [2,9]$} & \multicolumn{4}{c|}{Entire Wake: $x \in [2,50]$} \\ \hline \hline

\textbf{Term} &  \multicolumn{2}{c||}{$\overline{\omega}_t$} & \multicolumn{2}{c|||} {$\eta_t$} &  \multicolumn{2}{c||}{$\overline{\omega}_t$} & \multicolumn{2}{c|} {$\eta_t$} \\ \hline \hline

$\overline{\omega}$ & $\lambda_r$ & 0.0483 & $\lambda_i$ & 0.222  & $\lambda_r$ & –0.0269 & $\lambda_i$ &  0.499  \\ \hline
$\eta$ & $\lambda_i$ & 0.222  & $\lambda_r$ & 0.0483 & $\lambda_i$ &  0.499  & $\lambda_r$ & –0.0269  \\ \hline \hline

$\overline{\omega}^3$ & $\mu_{1,r}$  &  0.0583 & $\mu_{1,i}$ & –0.0732 & $\mu_{1,r}$ & –0.0154  & $\mu_{1,i}$ &  0.0787   \\ \hline
$\eta^3$ & $\mu_{1,i}$  & –0.0732 & $\mu_{1,r}$ &  0.0583 & $\mu_{1,i}$ &  0.0787  & $\mu_{1,r}$ & –0.0154   \\ \hline
$\overline{\omega}^2 \eta$ & $\mu_{2,r}$ &  0.154  & $\mu_{2,i}$ & –0.209  & $\mu_{2,r}$ & –0.0347  & $\mu_{2,i}$ & –0.00668   \\ \hline
$\overline{\omega} \eta^2$ & $\mu_{2,i}$ & –0.209  & $\mu_{2,r}$ &  0.154  & $\mu_{2,i}$ & –0.00668 & $\mu_{2,r}$ & –0.0347  \\ \hline \hline 

$\overline{\omega}_{x}$ & $\kappa_{r}$ & –0.613  & $\kappa_{i}$ &  0.0130 & $\kappa_{r}$ & –0.539   & $\kappa_{i}$ & 0.00715   \\ \hline
$\eta_{x}$ & $\kappa_{i}$ &  0.0130 & $\kappa_{r}$ & –0.613  & $\kappa_{i}$ &  0.00715 & $\kappa_{r}$ & –0.539 \\ \hline \hline 

$\overline{\omega}_{xx}$ & $\gamma_{r}$ & –0.00687 & $\gamma_{i}$ &  0.171   & $\gamma_{r}$ & –0.0102 & $\gamma_{i}$ & 0.231    \\ \hline
$\eta_{xx}$ & $\gamma_{i}$ &  0.171   & $\gamma_{r}$ & –0.00687 & $\gamma_{i}$ & 0.231  & $\gamma_{r}$ & –0.0102   \\ \hline \hline \hline 

\textbf{NRMSE}        &  \multicolumn{4}{c|||}{$6.7 \times 10^{-3}$} &  \multicolumn{4}{c|}{$3.2\times 10^{-2}$} \\ \hline
$\sigma$    &  \multicolumn{4}{c|||}{–0.389} &  \multicolumn{4}{c|}{0.0308} \\ \hline
$\beta$    &  \multicolumn{4}{c|||}{0.218} &  \multicolumn{4}{c|}{–0.0538} \\ \hline

\end{tabular}
\caption{Coefficients of generated sparse PDEs for coarse-grained vorticity in the form of Eqs. \eqref{eq:CGLE_2form_c}–\eqref{eq:CGLE_2form_d} in two different spatial domains: $x \in [2,9]$ (\textit{left}) and $x \in [2,50]$ (\textit{right}). For a given spatial domain's model, notice the symmetry in coefficient values, for example between the coefficient for $\overline{\omega}^3$ in the equation for $\overline{\omega}_t$ and the coefficient for $\eta^3$ in the equation for $\eta_t$. Such symmetries in terms are required by the form of the CGLE and discovered by SINDy naturally. The numerical solution of the near-wake model is shown in Figure \ref{fig:NumSoln} (\textit{bottom row}).}
\label{tb:PDEFIND_Coefs}
\end{center}
\end{table}

\begin{figure}[t!]{}
    \centering
    \includegraphics[width=7in]{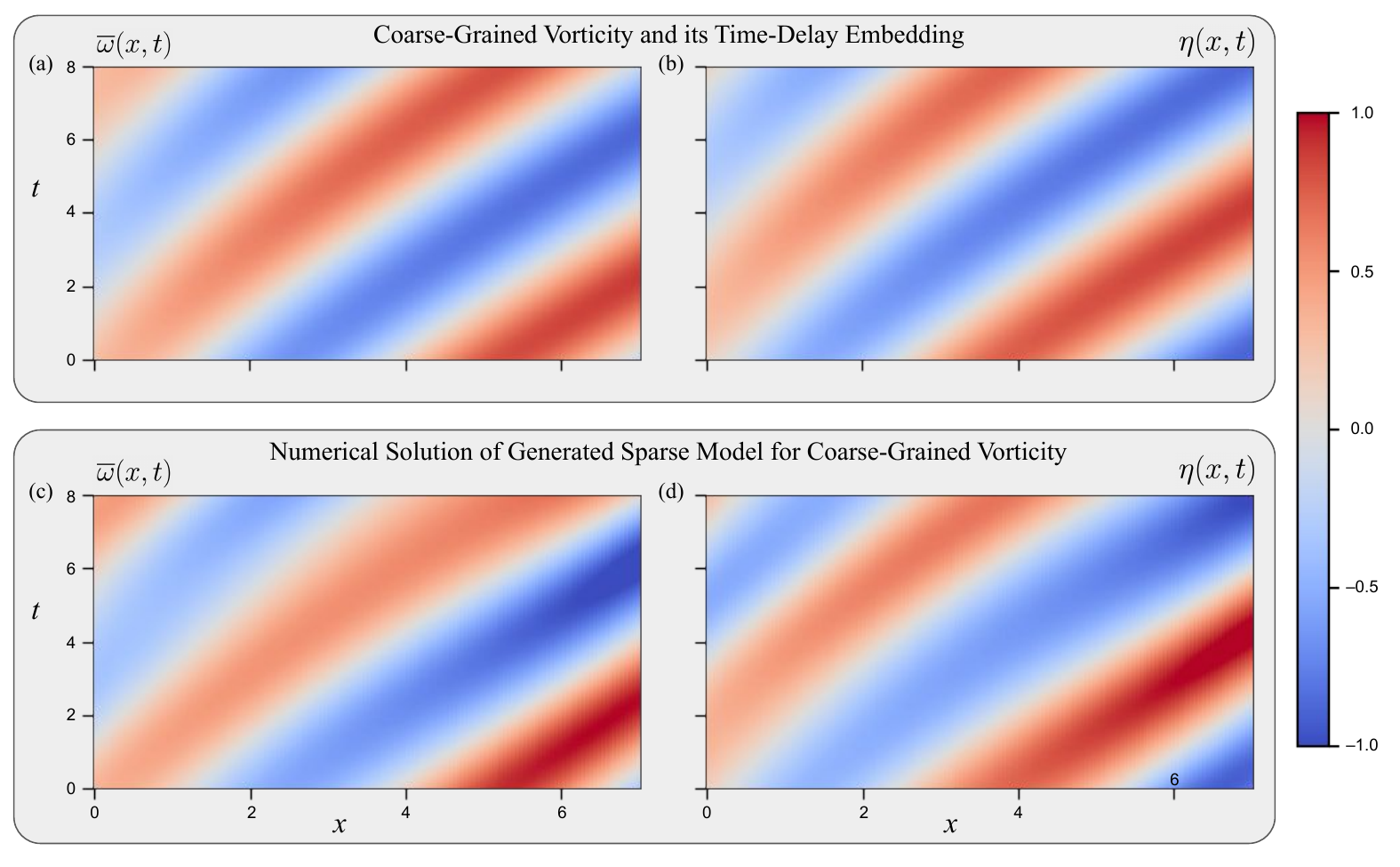}
    \caption{\textit{(a)-(b)}: The coarse-grained data $\mathcal{U} = (\overline{\omega}, \eta)$ used to generate the sparse PDE. \textit{(c)-(d)}: The pseudospectral numerical solution of the sparse model generated by SINDy, with coefficients given in Table \ref{tb:PDEFIND_Coefs}. Because of the periodic boundary conditions required for our pseudospectral method, the initial pulse will eventually propagate throughout the entire domain and begin interacting with itself, behavior which is clearly not reflected in the original data. To prevent this, the equations are solved on a large subdomain with regularizing hyperdiffusion in order to damp the self-interaction. Implemented in this way, there is striking similarity between the data and the numerical solution of the generated model. This close agreement validates the one-dimensional CGLE as a ROM of two-dimensional vortex shedding.}
    \label{fig:NumSoln}
\end{figure}

Having found a sparse system of PDEs that closely fits our data with minimal reconstruction error, we use a pseudospectral integration method to compute a numerical solution for the SINDy model generated from the near-wake. We focus on this model for now because of its lower error; local models for different subregions of the wake and global models for the entire wake will be explored more later. Figure \ref{fig:NumSoln} compares the original data $\mathcal{U}$ in the \textit{top row} to the numerical solution of the generated system of PDEs in the \textit{bottom row}. The numerical solution shows a close agreement with the data in the near-wake, reinforcing the validity of the CGLE as a model for our data and as a ROM for vortex shedding.

The numerical solution of our generated model closely matches the original data and lends great insight into its behavior, but some of its features are an approximation. Although convenient and simple to implement, a pseudospectral method supposes periodic boundaries which our original data clearly does not have. Our data originally comes from the vortex shedding wake behind a cylinder, and once coarse-grained, there is clearly strongly time-dependent behavior at the boundaries. In particular, the behavior of $\overline{\omega}$ near the surface of the cylinder is nearly discontinuous. This poses a challenge, because the model found by SINDy is generated without consideration of well-posedness and boundary conditions. 

Further, periodic boundary conditions allow our initial pulse to eventually self-interact once it has propagated the length of the domain. To overcome this, we embed the solution in a large spatial domain and investigate a spatiotemporal subdomain comparable to the original data, early enough in time that the initial pulse has not yet significantly interacted with itself. Although this strategy works well, it works better when we also include regularization in the form of hyperdiffusion terms (fourth-order spatial derivatives e.g. $\eta_{xxxx}$) in each equation with coefficient $R$. Hyperdiffusion serves to damp high frequency oscillations that result from self-interaction in a periodic domain. With distinct $\mu_1$ and $\mu_2$ and the inclusion of hyperdiffusion, our final equations then become the following:
\begin{subequations}
    \begin{align}
        \frac{\partial \overline{\omega}}{\partial t} = \lambda_r \overline{\omega} - \lambda_i \eta - \mu_{1,r} \overline{\omega}^3 + \mu_{2,i} \overline{\omega}^2 \eta - \mu_{2,r} \overline{\omega} \eta^2 + \mu_{1,i} \eta^3 + \kappa_r \overline{\omega}_{x} - \kappa_i \eta_{x} + \gamma_r \overline{\omega}_{xx} - \gamma_i \eta_{xx} + R \overline{\omega}_{xxxx}, \label{eq:CGLE_2form_e_HypDiff} \\
        \frac{\partial \eta}{\partial t} = \lambda_i \overline{\omega} + \lambda_r \eta - \mu_{1,i} \overline{\omega}^3 - \mu_{2,r} \overline{\omega}^2 \eta - \mu_{2,i} \overline{\omega} \eta^2 - \mu_{1,r} \eta^3 + \kappa_i \overline{\omega}_{x} + \kappa_r \eta_{x} + \gamma_i \overline{\omega}_{xx} + \gamma_r \eta_{xx} + R \eta_{xxxx}. \label{eq:CGLE_2form_f_HypDiff}
    \end{align}
\end{subequations}

While we use a large spatial domain in order to prevent self-interaction for as long as possible, too large of a domain requires infeasibly long calculation time. Hyperdiffusion with sufficiently large $R$ serves to damp the higher-order oscillations that begin once the signal begins self-interacting; however, we generally want a value of $R$ as low as possible as it will damp the main oscillations as well. For the numerical solution in the bottom row of Figure \ref{fig:NumSoln}, we use $R = 10^{-3}$. This value of $R$ was found to be optimal, with larger values of $R$ resulting in the same signal but with lower amplitude, and smaller values of $R$ resulting in quickly unstable solutions. Regularization in the form of hyperdiffusion is easy to include in the framework of a pseudospectral numerical solution and is highly effective. 

The analysis of this section has found close agreement between the original data, the model generated by SINDy, and the numerical solution of that model. As such, our method of coarse-graining, time-delay embedding, and model discovery with SINDy has resulted in a novel one-dimensional CGLE for two-dimensional cylinder vortex shedding, incorporating for the first time the streamwise coordinate into the Landau variable. In the following subsection, we will explore the spatial variation of local models found at different points in the wake.

\subsection{Parameter Tuning – Spatial Domain and Model Term Libraries}
In the previous subsection, we found a sparse model in the form of the CGLE to explain the dynamics of the coarse-grained vorticity in both the near-wake and the entire wake. Now we refine our method by varying and tuning certain parameters of our data-driven method. First we vary the spatial subdomain of interest and show how the error of the fitted model varies as the subdomain moves downstream. Then we expand the library of terms used to generate the model in order to better capture higher-order behavior in our data. We discuss each analysis in turn.

Regarding the spatial domain of the data, the initial work of \S~4.1 investigated both a specific flowfield subdomain of $x \in [2, 9]$  inspired by Rudy et al.~\cite{rudy_data-driven_2017}, and a much larger flowfield of $x \in [2, 50]$ to fully investigate the global wake dynamics. From this, it was clear that the data, the model coefficients, and the error are spatially dependent, so we next investigate how generated CGLE models vary depending on the location within the wake they are modeling. Specifically, for a subdomain of length $W_L$, SINDy generates a model from a specified library. The spatial subdomain has a midpoint $x_{mid}$ that then increments downstream, and the SINDy fit process is repeated. At each station within the wake, the NRMSE of the fitted model and its normal form parameters are computed. This entire process is then repeated for different values of $W_L$. In the limit $W_L \rightarrow 0$, the spatial coordinate of the coarse-grained data vanishes and our spatiotemporal data $\mathcal{U}(x,t)$ collapses to purely temporal data $\mathcal{U}(t;x_{mid})$ at that point $x_{mid}$. For this limiting case, we use SINDy to fit an ordinary differential equation in the form of the SLE (Eq. \eqref{eq:SLE_complex}), which mirrors the CGLE (Eqs. \eqref{eq:CGLE_2form_c}–\eqref{eq:CGLE_2form_d}) but without the spatial derivative terms. This is ``Library SL" and the full list of terms is shown in Appendix A.

Regarding the library of model terms, the initial work applied SINDy to the data with Library GL, a judiciously chosen library of candidate functions $\Theta$ with terms exactly matching those in the CGLE. Now we will allow for additional nonlinear terms in three forms: higher-order polynomial terms up to fifth order, products of the polynomial and derivative terms (e.g. $\eta^2 \eta_{xx}$), and combinations of those. These libraries are named for the total number of terms they feature, and respectively are referred to as ``GL$_{24}$", ``GL$_{49}$", and ``GL$_{104}$". For reference, the SL and GL Libraries have 9 and 13 terms. The full list of terms for all libraries are shown in Appendix A. These libraries contain many terms, and some of the larger libraries will have terms that contribute very little to the observed dynamics. Later we will tune the sparsity parameter $\lambda$ from Eq. \eqref{eq:PDEFIND_OptimizationProblem} in order to eliminate some terms from these larger libraries, leaving behind only those additional nonlinear terms that better explain the dynamics of the coarse-grained vorticity. The best sparse models will explain as much of the dynamics with as few terms as possible. For now, the $\text{NRMSE}$ of the nonsparse systems will provide a baseline to which we may later compare the sparse models.

\begin{figure}[t!]{}
    \centering
    \includegraphics[width=6.75in]{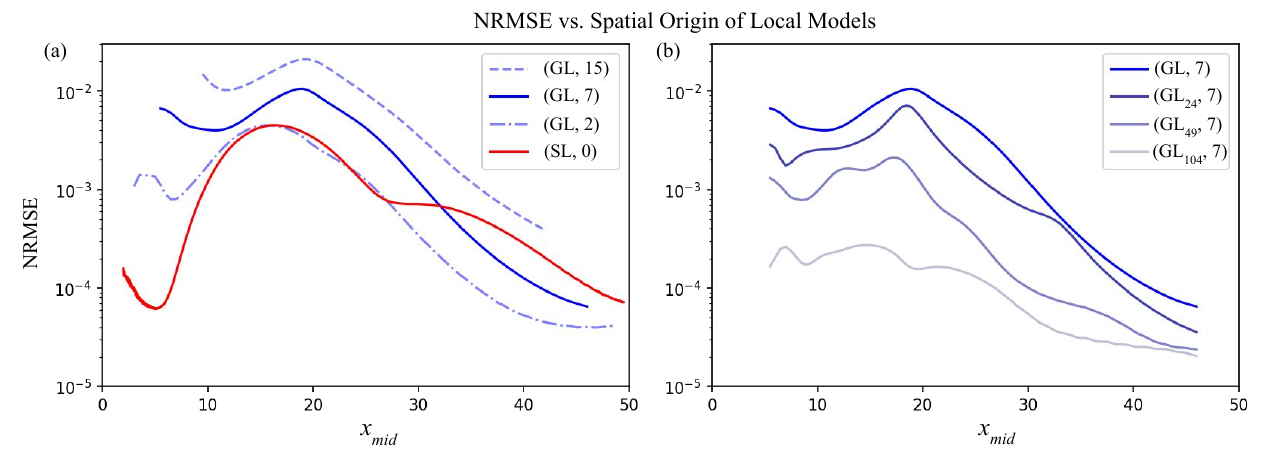}
    \caption{ \textit{(a)} The NRMSE of a generated local model vs. $x_{mid}$, the spatial midpoint of that model. A local CGLE model is generated from a subdomain of the data with midpoint $x_{mid}\in[2,50]$ and spatial length $W_L$. The NRMSE of this local model is found, $x_{mid}$ moves a short distance downstream, and the process repeats. In blue is the base case (GL,7), a function library with terms from the CGLE (Eqs. \eqref{eq:CGLE_2form_c}-\eqref{eq:CGLE_2form_d}) over a spatial subdomain with $W_L = 7$. In a lighter blue and with different linestyles are the NRMSE for cases with larger and smaller $W_L$, (GL,15) and (GL,2). They produce models with comparable errors, with the error generally decreasing for models on smaller subdomains. In red is the case (SL,0), where the spatial dimension of our training data vanishes, $W_L = 0$. In this case, we train using the SL Library and produce an ordinary differential equation at each spatial increment. 
    \textit{(b)} NRMSE vs. $x_{mid}$ for models trained with constant $W_L$ from libraries with additional terms beyond the GL Library. These libraries have the GL Library as their base, then either additional higher-order terms, nonlinear terms, or both. GL$_{24}$ has fourth- and fifth-order polynomial terms, GL$_{49}$ contains nonlinear differential terms, and GL$_{104}$ contains both. Their subscripts refer to the total number of terms in that library; the SL and GL Libraries have 9 and 13 terms, respectively. Consult Appendix A for a detailed list of all terms in each library. The base case of (GL, 7) in blue is the only curve in common between \textit{(a)} and \textit{(b)}.
    }
    \label{fig:NRMSE_vs_Subdomain_and_Library}
\end{figure}

As before, we quantify the goodness of fit of the model returned by SINDy with the $\text{NRMSE}$. Figure \ref{fig:NRMSE_vs_Subdomain_and_Library} shows how the local models' $\text{NRMSE}$ varies spatially in the wake, with parameterization by the spatial subdomain size and by the choice of library. Figure \ref{fig:NRMSE_vs_Subdomain_and_Library}(a) shows how the NRMSE from models generated with our initial GL Library varies with spatial location in the wake, plotted specifically against $x_{mid}$, the midpoint of the spatial domain. In dark blue is the base case using (GL,7). In a lighter shade of blue and with different linestyles are the NRMSE for cases with larger and smaller $W_L$, (GL,15) and (GL,2). In red is the case (SL,0), which does not incorporate any spatial derivative terms. Figure \ref{fig:NRMSE_vs_Subdomain_and_Library}(b) shows the spatial variation of NRMSE from models generated with equal subdomain sizes $W_L = 7$ and increasingly larger libraries ``GL$_{24}$", ``GL$_{49}$", and ``GL$_{104}$". The dark blue case of (GL,7) is identical on both plots of \ref{fig:NRMSE_vs_Subdomain_and_Library}.

In both Figures \ref{fig:NRMSE_vs_Subdomain_and_Library}(a)-(b), the NRMSE shows clear spatial dependence in for all $W_L$ and model library. The error is moderate in the near-wake, then grows to a global maximum at $x \approx 20$ in the mid-wake before a sharp decline in the far-wake. Indeed, the $\text{NRMSE}$ is 1-2 orders of magnitude lower in the far-wake than at its maximum in the mid-wake. In the mid-wake, the dynamics are characterized by the nonlinear growth and saturation of the SLE and CGLE and potentially by higher-order effects, but in the far-wake, the dynamics are primarily dominated by diffusion. Since the NRMSE is lower where diffusive dynamics dominate, this behavior is better explained by the terms in GL Library. Furthermore, in Figure \ref{fig:NRMSE_vs_Subdomain_and_Library}(a), the NRMSE generally decreases as $W_L$ decreases but variation with $x_{mid}$ remains similar, suggesting more accurate models can be fit to a more narrow spatial slice of the data. However, the limiting case of $W_L = 0$, which necessarily has no spatial component and hence no spatial derivatives in its model term library, has larger NRMSE in the far-wake region than does the base case of (GL,$W_L = 7$), reinforcing that derivative and specifically diffusive terms dominate the far-wake coarse-grained dynamics.

In Figure \ref{fig:NRMSE_vs_Subdomain_and_Library}(b), the $\text{NRMSE}$ of all local models improves as the library of candidate functions $\Theta$ becomes larger. Interestingly, the peak at $x \approx 20$ is much less pronounced for Library GL$_{104}$ (the largest library), though the error in the far-wake is still an order of magnitude lower. This suggests that the mid-wake region of the flowfield is more influenced by higher-order effects, hence when higher-order terms are included in the model, the error at the mid-wake greatly decreases. In the far-wake, where the dynamics are dominated by diffusion, the inclusion of additional higher-order terms helps the least. The minimum $\text{NRMSE}$ achieved anywhere across this parameter and hyperparameter tuning is $\approx 2\times10^{-5}$, occurring far downstream for a model generated from Library GL$_{104}$. This is over 2 magnitudes of an improvement in $\text{NRMSE}$ over the initial case (Table \ref{tb:PDEFIND_Coefs}). However, this has come at the cost of interpretability, as with zero sparsity ($\lambda=0$) this model has over 100 terms and cannot feasibly be numerically simulated; in Appendix B, we will tune our sparsity parameter $\lambda$ to remove those terms which have the least effect on the error.

\begin{figure}[t!]{}
    \centering
    \includegraphics[width=7in]{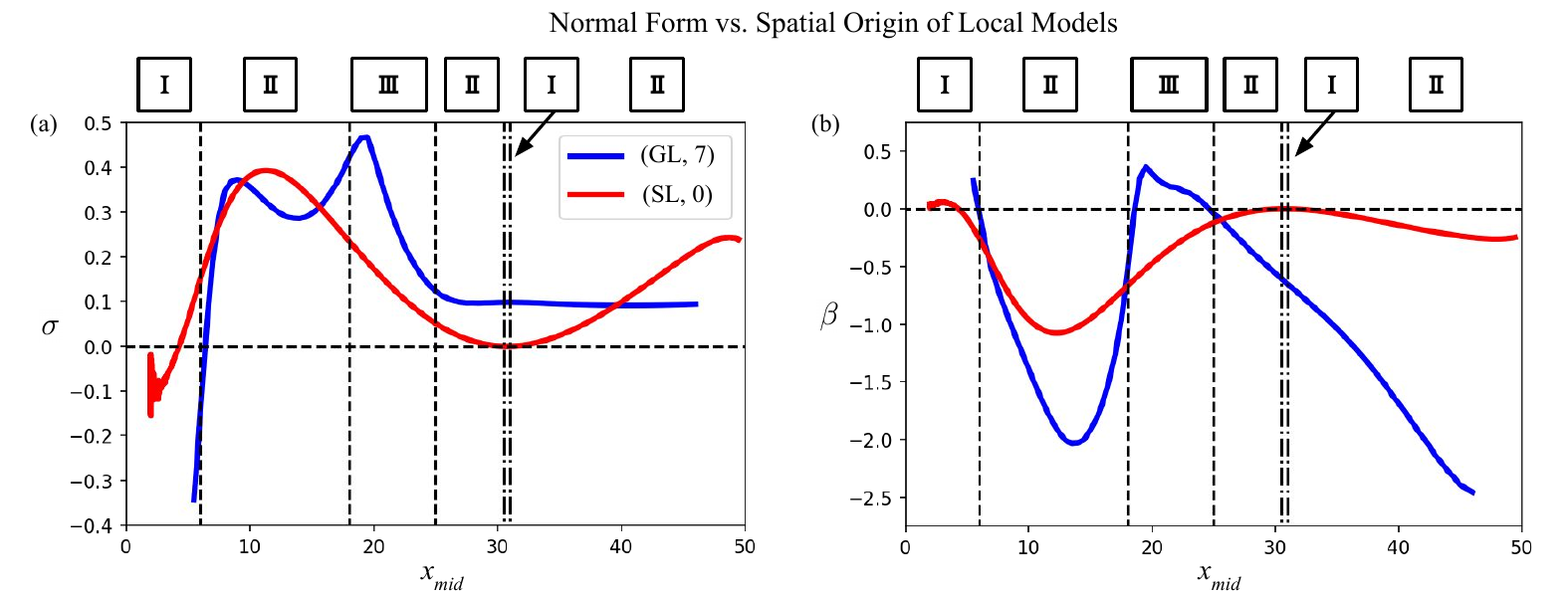}
    \caption{The normal form parameters $\sigma$ \textit{(a)}, and $\beta$ \textit{(b)} of a generated local model vs. $x_{mid}$, the spatial midpoint of that model. A local CGLE model is generated from a subdomain of the data with midpoint $x_{mid}\in[2,50]$ and spatial length $W_L$. The normal form parameters of this local model are found, $x_{mid}$ moves a short distance downstream, and the process repeats. In blue is the base case (GL,7), a function library with terms from the CGLE (Eqs. \eqref{eq:CGLE_2form_c}-\eqref{eq:CGLE_2form_d}) over a spatial subdomain with $W_L = 7$. In red is the case (SL,0), where the spatial dimension of our training data vanishes, $W_L = 0$. In this case, we train using the SL Library and produce an ordinary differential equation at each spatial increment. The normal form parameters $\sigma$ and $\beta$ are the coefficients of the linear and cubic terms of the normal form of the SLE (Eqn. \eqref{eq:Normal_Form}, and they characterize the systems stability and limit cycles. Three different regimes are denoted by I, II and III, where region I is $(\sigma < 0 , \beta > 0)$, region II is $(\sigma > 0 , \beta < 0)$, and region III is $(\sigma > 0 , \beta > 0)$. Consult Table \ref{tb:Hopf_Behavior} for further details. Most of the regions, delineated by $--$, occur for both GL and SL Library models. At $x \approx 30$, delineated by $-.-$, region I briefly returns for SL Library models only.}
    \label{fig:Sigma_and_Beta}
\end{figure}

The local models clearly vary spatially, hence by investigating the coefficients of the models in each subdomain and computing their normal form parameters $(\sigma, \beta)$, we can draw conclusions about the behavior of that local model and the spatial characteristics of the coarse-grained data. Figure \ref{fig:Sigma_and_Beta} shows $\sigma$ and $\beta$ vs. $x_{mid}$ for both the spatiotemporal ($W_L \neq 0$) CGLE models from Library GL and for purely temporal ($W_L = 0$) SLE models from Library SL. The normal form parameters were also computed for models (GL,15) and (GL,2) from Figure \ref{fig:NRMSE_vs_Subdomain_and_Library}(a), but were found to deviate only  marginally from the base case (GL,7), and hence are not shown. This suggests that $W_L$ does not have a strong effect on the normal form parameters of the local model, at least not until it vanishes for (SL,0).

The behavior of a Hopf bifurcation depends on the signs of $(\sigma, \beta)$; specifically, the sign of $\beta$ determines whether the fixed point is stable ($\beta<0$) or unstable ($\beta<0$), and the sign of $\sigma$ determines whether the nonlinear terms cause a stable limit cycle ($\sigma<0$) or unstable limit cycle ($\sigma>0$). As the spatial domain moves downstream, the local model transitions between several distinct regions of behavior, delineated in Figures \ref{fig:Sigma_and_Beta}(a)-(b). For each of these regions, Table \ref{tb:Hopf_Behavior} shows the normal form parameters $(\sigma, \beta)$, the stability of the equilibrium at the origin, the existence and stability of the limit cycle, and figures of the normal form phase portraits.

\begin{table}[ht]
    \centering
    \begin{tabular}{|m{3cm}||m{3cm}|m{3cm}|m{3cm}|}
        \hline
        & \textbf{Region I} & \textbf{Region II} & \textbf{Region III} \\ \hline \hline 
        
        Spatial Domain & $0 \lesssim x \lesssim 5$ \; and \;\;\;\;\;\;\;\;\; $x \approx 30$ & $5 \lesssim x \lesssim 18$ and \;\;\;\;\;\;\;\;\; $25 \lesssim x < +\infty$ & $18 \lesssim x \lesssim 25$  \\ \hline
        
        $\sigma$ & $\sigma<0$ & $\sigma>0$ & $\sigma>0$  \\ \hline
        
        $\beta$  & $\beta>0$  & $\beta<0$  & $\beta>0$    \\ \hline
        
        Origin Stability & Unstable & Asymptotically Stable & Unstable  \\ \hline
        Limit Cycle      & Stable   & Unstable              & None      \\ \hline
        Phase Portrait & \includegraphics[width=3cm,height=3.75cm]{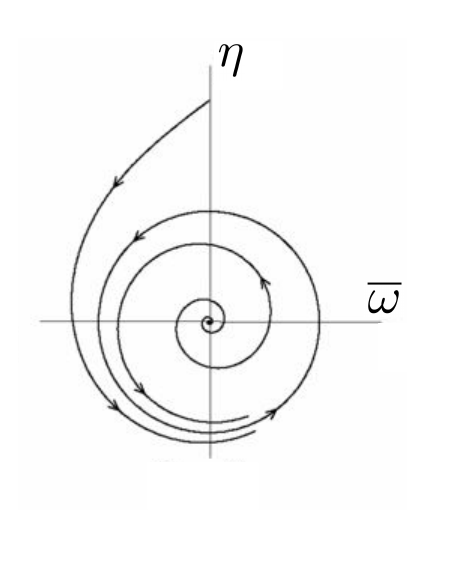} & \includegraphics[width=3cm,height=3.75cm]{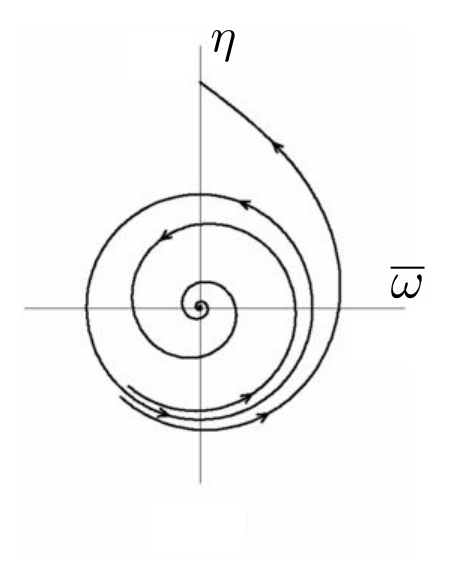} & \includegraphics[width=3cm,height=3.75cm]{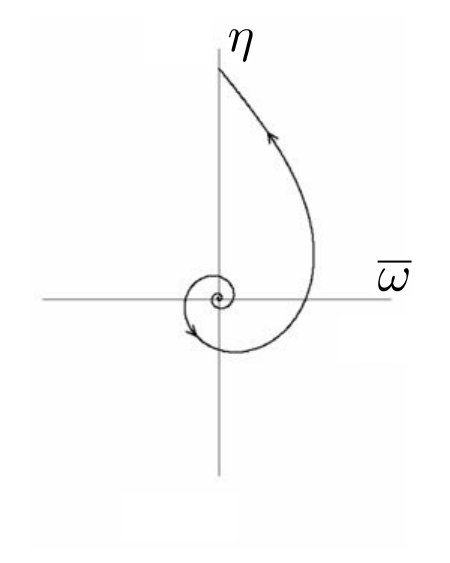} \\ \hline
    \end{tabular}
    \caption{The range of Hopf bifurcation behaviors and normal form parameter regimes present in the coarse-grained data from vortex shedding past a cylinder near the bifurcation point. From, Table \ref{tb:PDEFIND_Coefs}, recall that the global model for the entire domain $x \in [2,50]$ had normal form parameters $(\sigma > 0, \beta <0)$, matching Region II.}
    \label{tb:Hopf_Behavior}
\end{table}

Near the cylinder for $0 \lesssim x \lesssim 5$, Region I has normal form parameters ($\sigma<0$, $\beta>0$). Moving downstream, $\sigma$ and $\beta$ undergo a variety of sign and magnitude changes, transitioning through three unique sets of ($\sigma$, $\beta$). Throughout the wake, the normal form parameters from the GL Library and SL Library fits and generally agree with each other. While derivative terms are excluded from the SL Library, the nature of the data clearly suggests advective and diffusive behavior. In the absence of derivative terms, the values of the remaining coefficients compensate, and their changing values produce different $(\sigma, \beta)$. The main distinction is the differing behavior in the far wake, namely the $(SL,0)$ fits never experience Region III and instead experience a brief return of Region I at $x \approx 30$. Both regions feature $\beta>0$ and are distinguished by whether $\sigma<0$ (Region I) or $\sigma>0$ (Region III). For the (GL,7) model, $\beta>0$ for $18 \lesssim x \lesssim 25$ while $\sigma$ remains positive but decreasing to an asymptotic value. For the $(SL,0)$ model however, $\sigma$ becomes negative and $\beta$ simultaneously becomes positive, then both quickly switch back. Although difficult to determine visually, close inspection reveals that the bounds of these regions for both parameters coincide exactly, and both parameters are $\mathcal{O}(10^{-4})$. Recalling Figure \ref{fig:NRMSE_vs_Subdomain_and_Library}, the NRMSE produced by the model $(SL,0)$ sees a pause in its decline at $x \approx 30$ while other models' NRMSEs continue to decrease. This indicates that the fit from $(SL,0)$ in this region is relatively worse and the dynamics here may be due to higher-order effects not captured by the CGLE and not represented by its normal form parameters.

Further regarding the far-wake behavior, there is suggestive behavior indicative of long-range spatial variation in both parameters. While $\sigma$ reaches an asymptotic value in the far-wake for Library GL fits, $\sigma$ from Library SL briefly becomes negative in the far-wake before increasing and possibly attaining a secondary local maxima at $x \approx 50$. Regarding $\beta$, for all three fits the parameter decreases moving downstream after its maximum, but appears to begin increasing again far downstream at $x \approx 45$. Although the present spatial domain of analysis is insufficient to clearly capture the behavior, it appears that the local models' evolve on a large length scale and the normal forms have long-range spatial variation.

The spatial variation of normal forms may also suggest a connection with the wavemaker region, a region of local \textit{absolute} instability known to exist in the wake close to the cylinder, $0 \lesssim x \lesssim 4$~\cite{bagheri_input-output_2009,chomaz_global_2005}. In this region, perturbations grow and ultimately affect the entire domain. This near-wake region corresponds to Region I in Figure \ref{fig:Sigma_and_Beta}, where $\beta>0$ indicating that the fixed point is unstable and perturbations to this state grow. Sufficiently far downstream, the local flowfield is \textit{convectively} but not absolutely unstable, and correspondingly, $\beta<0$, signaling that perturbations from the stable fixed point decay, consistent with a flowfield in which the vorticity slowly decays away far downstream. The recurrence of regions of $\beta>0$ amid the downstream diffusion may again indicate longer-range behavior or a secondary wavemaker region.

\subsection{Heterogeneous Models}
\begin{figure}[t!]{}
    \centering
    \includegraphics[width=3.5in]{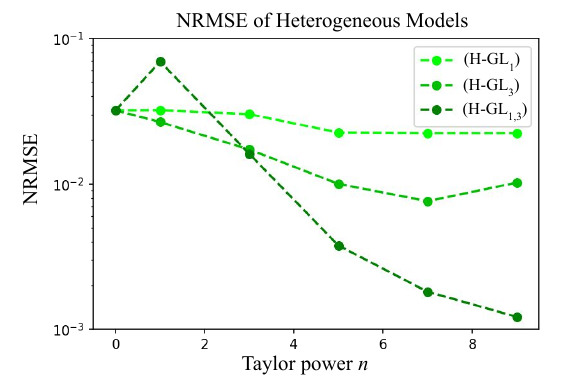}
    \caption{ 
    The NRMSE of a generated global heterogeneous model vs. $n$, the order of the Taylor polynomials in $x$ being used in that model. Taylor polynomials in $x$ as coefficients of the base GL Library terms directly allows for spatial variation in the global model, producing a heterogeneous CGLE model. Three different heterogeneous libraries are used, each with the GL Library as its base but with Taylor polynomials in $x$ of order $n$ fitted to either: just the linear terms (H-GL$_1$), just the cubic terms (H-GL$_3$), or both linear and cubic terms (H-GL$_{1,3}$). Appendix A shows the full list of terms in each library. }
    \label{fig:x_Coefs_and_TaylorPolyns}
\end{figure}

The analysis of the previous section explored generated local models' spatial behaviors and variations. In this section, we further iterate on the form of the model and minimize the NRMSE by fitting a global heterogeneous CGLE to the data of the entire wake $x \in [2,50]$. This is done by introducing $x-$dependent coefficients to replace constant coefficients on the model terms. By explicitly allowing for spatial variation in the coefficients of the terms, we find a global model for the entire wake that captures the spatially changing dynamics. The general form of our dynamical system becomes:
\begin{equation} \label{eq:PDEFIND_DynSyst_Het}
\mathcal{U}_{t} =  \mathcal{N}(g_1(x) \cdot f_1(\mathcal{U}), ... , g_i(x) \cdot  f_i(\mathcal{U}), ...  ),
\end{equation}
where our data $\mathcal{U} = [\overline{\omega} , \eta]$ is again our coarse-grained vorticity and its time-delay embedding. First, we will try using $g_i(x)$ as polynomials in $x$ up to a specified order as an approximation to the Taylor series of the true, unknown coefficient functions for each term in the Library GL. In Figure \ref{fig:x_Coefs_and_TaylorPolyns}(a), the NRMSE of a generated model is plotted against the order $n$ of the Taylor polynomials. The different libraries are:
\begin{itemize}
    \item H-GL$_{1}$ – Taylor series in $x$ on the linear terms of the GL library
    \item H-GL$_{3}$ – Taylor series in $x$ on the cubic terms of the GL library
    \item H-GL$_{1,3}$ – Taylor series in $x$ on both the linear and cubic terms of the GL library
\end{itemize}
At $n=0$, the coefficients are constant and we reproduce the initial case with Library GL on the global wake. The NRMSE then improves with Taylor polynomials on the linear terms, more with Taylor polynomials on the cubic terms, and even more with Taylor polynomials on both the linear and cubic terms. The NRMSE generally improves with higher order $n$ of the Taylor polynomials, although improvements generally diminish around $n=7$, after which the NRMSE of the generated model experiences only marginal improvements. At $n=9$, Library H-GL$_{1,3}$ achieves a minimum NRMSE of $1.2\times10^{-3}$, an order of magnitude improvement over the initial case (NRMSE = $3.2\times10^{-2}$). The heterogeneous system's improvement of the global model's NRMSE further reinforces that the vortex shedding wake and its ROMs have spatially-dependent behavior that a homogeneous global model fails to capture.

\section{Conclusion}

In this work, we developed a novel one-dimensional Ginzburg-Landau model for two-dimensional cylinder vortex shedding, incorporating the streamwise coordinate of the flow as the spatial coordinate of the Landau model. After simulating flow past a cylinder at $\text{Re} = 50$, the two-dimensional vorticity field $\omega(x,y,t)$ is first \textit{coarse-grained}, producing $\overline{\omega}(x,t)$. This signal is then \textit{time-delay embedded}, producing the system $\mathcal{U} = [\overline{\omega} , \eta](x,t)$. SINDy is used to fit a model to the observed dynamics. Using a library of terms modeled after the CGLE, the data is modeled and reconstructed with minimal error. The numerical solution of the generated model reproduced the near-wake coarse-grained dynamics almost exactly. 

From this initial analysis, it was clear that the data and models were spatially dependent, so we next investigated how the generated CGLE models varied depending on the location within the wake they were modeling. By fitting a CGLE model to a given spatial subdomain and then sliding downstream, we developed smoothly varying local models throughout the entire wake. At each station within the wake, the fitted model's NRMSE and normal form parameters ($\sigma, \beta$) were computed. Figures \ref{fig:NRMSE_vs_Subdomain_and_Library} and \ref{fig:Sigma_and_Beta} showed the spatial variation in NRMSE and ($\sigma, \beta$) across different model libraries. These findings clearly showed how the properties of the local models divided the entire wake into distinct subregions. Furthermore, libraries with higher-order terms beyond the base CGLE terms uniformly improved the NRMSE of local models, but this improvement was most pronounced in the mid-wake region and least evident in the far-wake region. From this, it was clear that the near- and mid-wake regions were characterized by higher-order behavior that was less explained by the CGLE models, while the far-wake was dominated by diffusive dynamics. Once sparsity is introduced (Appendix B), using larger libraries of higher-order terms provided only limited reduction in error; at their core, the coarse-grained dynamics are modeled by the CGLE. Finally, the behavior of local and global models was contrasted by fitting a global heterogeneous model to the coarse-grained data of the entire wake. This was done by applying Taylor polynomials in $x$ of some order $n$ to the linear and cubic terms of the CGLE. The best global heterogeneous model improved upon the NRMSE of a global homogeneous model by over an order of magnitude, further confirming that the local behavior of the wake varies downstream.

These findings validate the CGLE as a ROM for two-dimensional cylinder vortex shedding. They join a host experimental, numerical, and analytic work which has previously investigated the Stuart-Landau and Ginzburg-Landau equations as ROMs for two- and three-dimensional cylinder vortex shedding~\cite{sreenivasan_hopf_1987,dusek_numerical_1994,albarede_modelisation_1990,sipp_global_2007,provansal_benard-von_1987,le_dizes_weakly_1991}. In particular, the CGLE had previously only been associated with three-dimensional cylinder vortex shedding, incorporating the spanwise coordinate of the flow as the spatial coordinate the CGLE~\cite{albarede_modelisation_1990,le_dizes_weakly_1991}. Our work uniquely incorporates the streamwise coordinate from the flowfield as the spatial coordinate in our one-dimensional CGLE, which had seen only limited prior attempts~\cite{chiffaudel_nonlinear_1992,park_model_1992}. The incorporation the streamwise coordinate into the CGLE model for cylinder vortex shedding has helped shed light on the changes in stability and behavior of downstream local models.

A further contribution of this work is the justification of time-delay embedding as an appropriate method for discovering a multidimensional system from a single data signal. The SLE and CGLE may be presented as either a single complex differential equation or a pair of coupled, real differential equations, hence two (real) data signals are required in order to discover the SLE and CGLE with SINDy. Whereas in~\cite{loiseau_sparse_2018} the coefficient of lift $C_l(t)$ is time-differentiated in order to generate a second oscillatory, out-of-phase signal, we time-delay embed our data. In the case of $C_l(t)$, time-delay embedding generates similar results as time-differentiating: both cases produce an SLE model with low NRMSE and whose numerical solution closely matches the behavior of the original signal. In the case of $\overline{\omega}(x,t)$, time-differentiating introduces additional higher-order behavior less explainable by a sparse model, whereas time-delay embedding does not. Time-differentiation also introduces and amplifies numerical errors, whereas time-delay embedding does not.

Future work includes increasing the value of the control parameter $\text{Re}$ past the bifurcation and using the recent parameterized SINDy~\cite{nicolaou_data-driven_2023} to parameterize the local ROMs vs. $\text{Re}$. The SLE and CGLE only describe the Hopf bifurcation when the control parameter is in the vicinity of the bifurcation point~\cite{kuznetsov_elements_1998}, hence a similar study of local models but with $\text{Re} >> 47$ may help explain the effect of increasing $\text{Re}$ on the cylinder vortex shedding wake. Our methods may also help elucidate the normal forms of other bifurcations experienced in vortex shedding flows, for example those experienced in three-dimensional cylinder flow~\cite{barkley_bifurcation_2000}, in two-dimensional rotating cylinder flow~\cite{sierra_bifurcation_2020}, or in the flow over the fluidic pinball~\cite{deng_route_2018, deng_low-order_2020}. Common in biological applications, some vortex streets do not present as rows of alternating vortices but instead may shed multiple large or small vortices at once~\cite{kanso_locomotion_2005, colvert_classifying_2018}; coarse-graining such data would result in rich spatiotemporal signals, prime candidates for our method of data-driven model discovery.

A final extension of this work may be towards feedback control systems. The SINDy method has been applied to controlled systems with great success~\cite{kaiser_sparse_2018}, and control and suppression of vortex shedding, particularly vortex shedding past a cylinder, is a well-researched topic~\cite{sumer_hydrodynamics_2006,rashidi_vortex_2016}. There are many different strategies for vortex shedding control, from passive modifications to the body~\cite{larsen_storebaelt_2000,strykowski_formation_1990} to active actuation of control surfaces~\cite{fujisawa_feedback_2001}. Analysis of the coarse-grained wake dynamics under different control strategies may help identify the extent to which these control strategies are successful and what regions of the wake experience the greatest vortex suppression effects.

\section*{Acknowlegments}
The authors acknowledge support from the National Science Foundation AI Institute in Dynamic Systems
(grant number 2112085), the US Air Force Office of Scientific Research (FA9550-21-1-0178), and The Boeing Company.

 \begin{spacing}{.87}
 \setlength{\bibsep}{5.5pt}

 \end{spacing}

\appendix
\section{Libraries of Candidate Functions}

In this work, SINDy is used to find a dynamical system modeling given input data. The input data $\mathcal{U} = [\overline{\omega} , \eta]$ consists of two spatiotemporal signals: the coarse-grained vorticity and its time-delay embedding. Thus, the dynamical system will consist of two differential equations for $\overline{\omega}_t$ and $\eta_t$. The matrix of candidate functions $\Theta(u,v)$ contains the functions possible in \textit{both} equations, where $u$ and $v$ are our signals $\overline{\omega}$ and $ \eta $. \\~\\

\textit{SL} Library: \textit{Stuart-Landau equation} (SLE). Polynomials up to 3rd order. \\
$ \Theta = [ \; u, v, u^2, uv, v^2, u^3, u^2v, uv^2, v^3 \; ]   $ \\

\textit{GL} Library: \textit{Complex Ginzburg-Landau equation} (CGLE). Polynomials up to 3rd order and derivatives up to 2nd order. \\
$ \Theta = [ \; u, v, u^2, uv, v^2, u^3, u^2v, uv^2, v^3, u_x, u_{xx}, v_x, v_{xx} \; ]   $ \\

\textit{GL$_{24}$} Library:  polynomials up to 5th order and derivatives up to 2nd order. \\
$ \Theta = [ \; u, v, u^2, uv, v^2, u^3, u^2v, uv^2, v^3, u^4, u^3v, u^2v^2, uv^3, v^4, u^5, u^4v, u^3v^2, u^2v^3, uv^4, v^5, u_x, u_{xx}, v_x, v_{xx} \; ] 
  $ \\

\textit{GL$_{49}$} Library: polynomials up to 3rd order, derivatives up to 2nd order, and cross-terms thereof. \\
$\Theta = [ \; u, v, u^2, uv, v^2, u^3, u^2v, uv^2, v^3, u_x, u_{xx}, v_x, v_{xx}, ...     $ \\
$ ... \; u_x \cdot (u, v, u^2, uv, v^2, u^3, u^2v, uv^2, v^3), \; ...     $ \\
$ ... \; u_{xx} \cdot (u, v, u^2, uv, v^2, u^3, u^2v, uv^2, v^3), \; ...     $ \\
$ ... \; v_x \cdot (u, v, u^2, uv, v^2, u^3, u^2v, uv^2, v^3), \; ...     $ \\
$ ... \; v_{xx} \cdot (u, v, u^2, uv, v^2, u^3, u^2v, uv^2, v^3) \; ]   $ \\

\textit{GL$_{104}$} Library: polynomials up to 5th order, derivatives up to 2nd order, and cross-terms thereof. \\
$\Theta = [ \; u, v, u^2, uv, v^2, u^3, u^2v, uv^2, v^3, u^4, u^3v, u^2v^2, uv^3, v^4, u^5, u^4v, u^3v^2, u^2v^3, uv^4, v^5, u_x, u_{xx}, v_x, v_{xx}, ...     $ \\
$ ... \; u_x \cdot (u, v, u^2, uv, v^2, u^3, u^2v, uv^2, v^3, u^4, u^3v, u^2v^2, uv^3, v^4, u^5, u^4v, u^3v^2, u^2v^3, uv^4, v^5), \; ...     $ \\
$ ... \; u_{xx} \cdot (u, v, u^2, uv, v^2, u^3, u^2v, uv^2, v^3, u^4, u^3v, u^2v^2, uv^3, v^4, u^5, u^4v, u^3v^2, u^2v^3, uv^4, v^5), \; ...     $ \\
$ ... \; v_x \cdot (u, v, u^2, uv, v^2, u^3, u^2v, uv^2, v^3, u^4, u^3v, u^2v^2, uv^3, v^4, u^5, u^4v, u^3v^2, u^2v^3, uv^4, v^5), \; ...     $ \\
$ ... \; v_{xx} \cdot (u, v, u^2, uv, v^2, u^3, u^2v, uv^2, v^3, u^4, u^3v, u^2v^2, uv^3, v^4, u^5, u^4v, u^3v^2, u^2v^3, uv^4, v^5) \; ]   $ \\

\textit{H-GL$_1$} Library: GL library with Taylor polynomials in $x$ up to $n$th order on the linear terms. \\
$ \Theta = [ \; (x + ... + x^n)u, (x + ... + x^n)v, ... \\
u^2, uv, v^2, u^3, u^2v, uv^2, v^3, u_x, u_{xx}, v_x, v_{xx} \; ]   $ \\

\textit{H-GL$_3$} Library: GL library with Taylor polynomials in $x$ up to $n$th order on the cubic terms. \\
$ \Theta = [ \; u, v, u^2, uv, v^2, ... \\
(x + ... + x^n)u^3, (x + ... + x^n)u^2v, (x + ... + x^n)uv^2, (x + ... + x^n)v^3, ... \\
u_x, u_{xx}, v_x, v_{xx} \; ]   $ \\

\textit{H-GL$_{1,3}$} Library: GL library with Taylor polynomials in $x$ up to $n$th order on both the linear and cubic terms. \\
$ \Theta = [ \; (x + ... + x^n)u, (x + ... + x^n)v, ... \\
u^2, uv, v^2, ... \\
(x + ... + x^n)u^3, (x + ... + x^n)u^2v, (x + ... + x^n)uv^2, (x + ... + x^n)v^3, ... \\
u_x, u_{xx}, v_x, v_{xx} \; ]   $ \\

\newpage
\section{Sparse Approximations of the CGLE}

\begin{figure}[t!]{}
    \centering
    \includegraphics[width=7in]{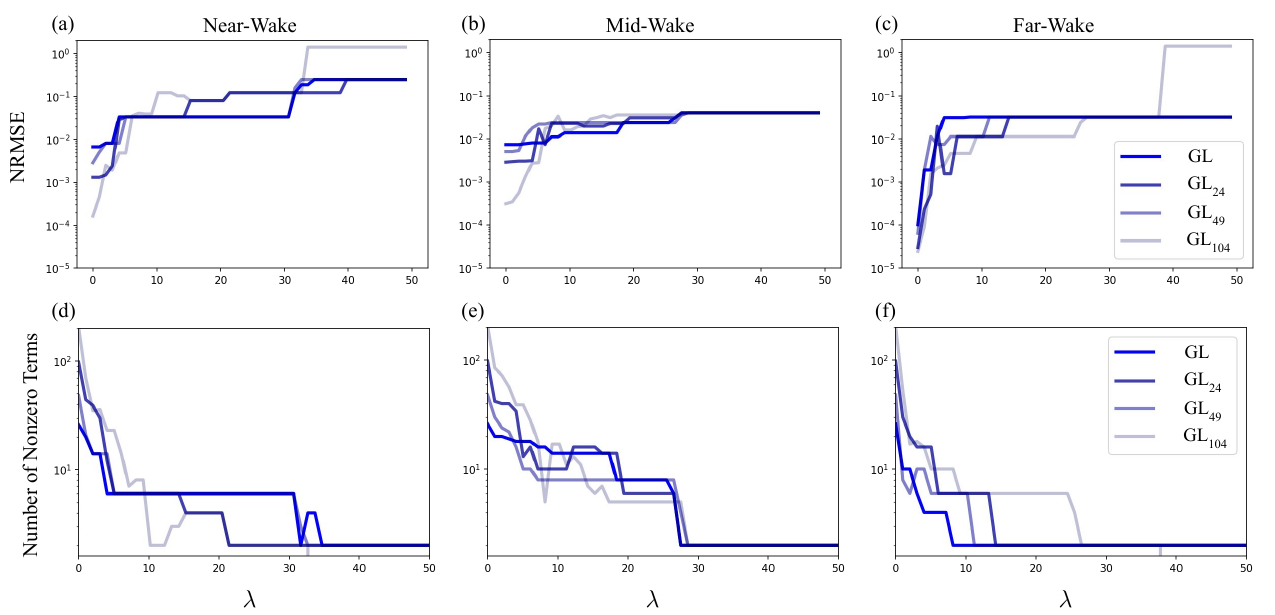}
    \caption{Models with increasing sparsity, controlled by the parameter $\lambda$, are generated for three different subregions of the wake; terms come from Library GL (in dark blue), which contains the standard CGLE terms. In lightening shades of blue are Libraries GL$_{24}$, GL$_{49}$, and GL$_{104}$, which contain the base terms from Library GL and various additional nonlinear terms. Their subscript denotes the total number of terms in their library, hence they get increasingly large. Consult Appendix A for a full list of terms. The NRMSE of generated models is shown vs. $\lambda$ in subdomains \textit{(a)} $x \in [2,9]$, \textit{(b)} $x \in [10,19]$, and \textit{(c)} $x \in [38,45]$, respectively the near-, mid-, and far-wakes. The number of nonzero terms of the generated models is shown vs. $\lambda$ in \textit{(d)}, \textit{(e)}, and \textit{(f)}, for the same subdomains. As expected, at $\lambda = 0$ the NRMSE is smallest for the libraries with the most terms, but this is least pronounced in the far wake \textit{(c)}. As $\lambda$ is increased, terms are dropped from the generated model, beginning with terms that have the smallest contribution to the overall dynamics. This necessarily increases the NRMSE, but retains the terms which most explain the observed dynamics. Increasing the sparsity is necessary when using Libraries GL$_{24}$, GL$_{49}$, and GL$_{104}$, as these libraries contain far more terms than the standard GL Library. Some libraries coincide and plateau for ranges of $\lambda$, particularly when $\lambda$ is large enough that many or most terms have been eliminated; these are models so sparse that they no longer capture the original dynamics.}
    \label{fig:NMSE_vs_Sparsity_Subdomains}
\end{figure}

In this section, we realize the full potential of SINDy by iterating upon the sparsity parameter $\lambda$. In previous sections, models with extremely low error were produced, but they have too many terms to feasibly implement numerically and solve. By tuning $\lambda$, SINDy will retain only those terms which most explain the observed dynamics, with larger $\lambda$ eliminating more terms. We focus on 3 equally-sized subdomains of the wake: the near-, mid-, and far-wakes, encompassing $x \in [2,9]$, $x \in [10,19]$, and $x \in [38,45]$, respectively. As bases for our sparse models, we use Libraries GL and its extensions GL$_{24}$, GL$_{49}$, and GL$_{104}$. We exclude heterogeneous libraries from this process as we seek sparse, interpretable equations that can be solved with our pseudospectral numerical method from \S~4.1. For each subdomain, Figures \ref{fig:NMSE_vs_Sparsity_Subdomains}(a)-(c) show NRMSE vs. $\lambda$ for each library and Figures \ref{fig:NMSE_vs_Sparsity_Subdomains}(d)-(f) show the number of nonzero terms in the model vs. $\lambda$.

The \textit{top row} of Figure \ref{fig:NMSE_vs_Sparsity_Subdomains} shows the NRMSE vs. $\lambda$ for each of the four libraries across the three wake subregions. The \textit{bottom row} of Figure \ref{fig:NRMSE_vs_Subdomain_and_Library} shows the number of nonzero terms of the generate models vs. $\lambda$. As expected, the NRMSE is lowest for each library when $\lambda = 0$. At this value, Library GL$_{104}$ has the lowest NRMSE, although this effect is least noticeable for the far-wake region. This suggests the dynamics of the near- and mid-wake regions are better explained by the additional nonlinear terms than are the far-wake dynamics, which are known to be dominated by diffusion. These findings align with those of Figure \ref{fig:NRMSE_vs_Subdomain_and_Library}, which shows generally similar dependence of the NRMSE on spatial region and model complexity. 
For each library and subdomain, as sparsity increases, the numbers of nonzero terms decrease and the NRMSEs increase from their nonsparse minimum values, with some of the curves reaching various plateaus. Once $\lambda$ has been increased large enough to eliminate \textbf{all} terms from a given model, that NRMSE reaches its maximum value of $\sqrt{2}$. For a given library, the errors are comparable in the near- and mid-wakes and an order of magnitude lower in the far-wake, reinforcing that the diffusion-dominated dynamics of the far-wake are most easily explained.

Regarding the plateaus, the models produced here are relatively stable with respect to the sparsity, as opposed to regions where the model produced (and hence NRMSE) changes with even small changes in the sparsity. Additionally, for a given region, the observed plateaus generally overlap between the different libraries. While Libraries GL$_{24}$, GL$_{49}$, and GL$_{104}$ outperform Library GL at $\lambda = 0$, once any sparsity is introduced, models from these libraries generally either underperform or perform no better than models from Library GL. Further, the higher-order terms in Libraries GL$_{24}$, GL$_{49}$, and GL$_{104}$ are among the first terms dropped as $\lambda$ increases, and the terms retained are either the standard CGLE terms from Library GL, or a higher-order version (e.g. $u_{xx}$ and $v_{xx}$ from Library GL vs. $u^2 u_{xx}$ and $u^2 v_{xx}$ from Library GL$_{49}$). Hence while larger libraries do improve the error for nonsparse sparse models, at their core the coarse-grained dynamics are still modeled very well by the CGLE, even with with only a subset of terms. When interpretability and solvability of the model are desired and sparsity is introduced, the higher-order terms vanish and there is no benefit to using Libraries GL$_{24}$, GL$_{49}$, and GL$_{104}$ over Library GL. Altogether, these findings support our development of a one-dimensional Ginzburg-Landau model as a ROM for two-dimensional cylinder vortex shedding.

\end{document}